\documentclass[onecolumn]{emulateapj}
\usepackage{longtable}
\usepackage{graphicx}
\usepackage{graphics}
\usepackage{color}
\usepackage{epsfig}
\usepackage{rotating}
\usepackage{subfigure}
\usepackage{float}
\usepackage{url}
\usepackage{hyperref}
\usepackage{breakurl}

\shortauthors{}
\shorttitle{}

\begin{document}

\title{\emph{XMM-Newton} Large Program on SN1006 - I: Methods and Initial Results of Spatially-Resolved Spectroscopy}

\author{Jiang-Tao Li\altaffilmark{1,5}, Anne Decourchelle\altaffilmark{1}, Marco Miceli\altaffilmark{2,4}, Jacco Vink\altaffilmark{3}, Fabrizio Bocchino\altaffilmark{4}} 

\altaffiltext{1}{Service d'Astrophysique, CEA Saclay, 91191 Gif-sur-Yvette Cedex, France} 

\altaffiltext{2}{Dipartimento di Fisica \& Chimica, Universit$\rm\grave{a}$ di Palermo, Piazza del Parlamento 1, I-90134 Palermo, Italy}

\altaffiltext{3}{Anton Pannekoek Institute/GRAPPA, University of Amsterdam, PO Box 94249, 1090 GE Amsterdam, The Netherlands}

\altaffiltext{4}{INAF-Osservatorio Astronomico di Palermo, Piazza del Parlamento, 90134 Palermo, Italy} 

\altaffiltext{5}{Department of Astronomy, University of Michigan, 311 West Hall, 1085 S. University Ave, Ann Arbor, MI, 48109-1107, U.S.A.}

\nonumber

\begin{abstract}
Based on our newly developed methods and the \emph{XMM-Newton} large program of SN1006, we extract and analyze the spectra from 3596 tessellated regions of this SNR each with 0.3-8~keV counts $>10^4$. For the first time, we map out multiple physical parameters, such as the temperature ($kT$), electron density ($n_e$), ionization parameter ($n_et$), ionization age ($t_{ion}$), metal abundances, as well as the radio-to-X-ray slope ($\alpha$) and cutoff frequency ($\nu_{cutoff}$) of the synchrotron emission. We construct probability distribution functions of $kT$ and $n_et$, and model them with several Gaussians, in order to characterize the average thermal and ionization states of such an extended source. We construct equivalent width (EW) maps based on continuum interpolation with the spectral model of each regions. We then compare the EW maps of \ion{O}{7}, \ion{O}{8}, \ion{O}{7}~K$\delta-\zeta$, Ne, Mg, \ion{Si}{13}, \ion{Si}{14}, and S lines constructed with this method to those constructed with linear interpolation. We further extract spectra from larger regions to confirm the features revealed by parameter and EW maps, which are often not directly detectable on X-ray intensity images. For example, O abundance is consistent with solar across the SNR, except for a low-abundance hole in the center. This ``O Hole'' has enhanced \ion{O}{7} K$\delta-\zeta$ and Fe emissions, indicating recently reverse shocked ejecta, but also has the highest $n_et$, indicating forward shocked ISM. Therefore, a multi-temperature model is needed to decompose these components. The asymmetric metal distributions suggest there is either an asymmetric explosion of the SN or an asymmetric distribution of the ISM.
\end{abstract}


\section{Introduction}\label{PaperIsec:Introduction}

In young supernova remnants (SNRs), both shocked thermal plasma emission and non-thermal emission from accelerated particles typically peak in the 0.5-10~keV X-ray band (e.g., \citealt{Reynolds08,Vink12,Slane14}). The X-ray properties of these emission components often show clear spatial variations. Therefore, spatially resolved X-ray observations of young SNRs play important roles in our understanding of many scientific issues, such as the explosion mechanisms and progenitors of supernova (SN), the microphysics involved in particle acceleration and magnetic field amplification, the heating of electrons and ions at the shock, as well as the abundances and the distribution of fresh nucleosynthesis products (see \citealt{Vink12} for a recent review).

Current facilities (in particular, X-ray CCDs) on board space X-ray telescopes such as \emph{Chandra} and \emph{XMM-Newton} have the ability to simultaneously record the spatial and spectral information of the collected photons. They could thus be used for spatially resolved spectroscopy analysis to study the spatial distribution of X-ray properties across young SNRs. However, methods commonly used in X-ray data analysis, such as spectral analysis of individual interesting regions (e.g., \citealt{Chen08}) and equivalent width (EW) map of strong emission lines (e.g., \citealt{Hwang00}), usually do not make full use of the information contained in the X-ray CCD data. What we want, and are often contained in the X-ray data, are the spatial distributions of various physical properties of the thermal and non-thermal X-ray emission across the SNRs.

In this paper, we will introduce new techniques to conduct spatially resolved spectroscopy analysis. The basic idea is to map out the spectral parameters in small tessellated regions which individually contain enough photons for spectral analysis. Such techniques have initially been developed for optical integral field observations \citep{Cappellari03}, and have later been applied to X-ray observations \citep{Diehl06,Broos10}, especially in resolving interesting features in the temperature/metallicity maps of massive galaxy clusters (e.g., \citealt{Randall08}) or interacting galaxies (e.g., \citealt{HodgesKluck12}). SNRs typically have more complicated X-ray spectra which often include ionization non-equilibrium thermal plasma (e.g., \citealt{Vink12,Slane14}) and/or non-thermal emission with varying spectral shapes (e.g., \citealt{Miceli14}). It is therefore more difficult to automatically decompose different spectral components and model the spectra extracted from a large number of tessellated regions. Furthermore, existing spatially resolved spectroscopy analysis of SNRs often adopt \emph{equal-sized} (e.g., \citealt{Lu00,Lopez13}) or adaptively-binned \emph{box-shaped} meshes (e.g., \citealt{CassamChenai04}), which are less efficient in resolving fine structures on physical parameter maps.

The remnant of the supernova AD1006 (SN1006) is one of the few historical SNRs with human records of its exact birth date (e.g., \citealt{Stephenson10}). It is widely accepted that SN1006 is the remnant of a Type~Ia supernova event, based on its high Galactic latitude location [$b=14.6^\circ$, apparently isolated from any star formation regions], the lack of any visible central compact sources (e.g., \citealt{Pye81,Jones89,Burleigh00}), the detection of iron absorption lines from the ultraviolet (UV) spectra of background sources (e.g., \citealt{Fesen88,Wu93}), the metal abundances inferred from the soft X-ray emission lines \citep{Koyama08,Uchida13}, the lack of time variation due to the clumpy of the surrounding medium \citep{Katsuda10}, and the historical records that it remains visible for several years (e.g., \citealt{Stephenson10}). Proper motion measurements in optical (e.g., \citealt{Long88}), radio (e.g., \citealt{Moffett93}), or X-ray \citep{Winkler14}, combined with the shock velocity as inferred from the measurements of the width and ratio of some optical/UV emission lines (e.g., \citealt{Kirshner87,Laming96,Ghavamian02}), or the expanding velocity inferred from the broadening of some UV absorption lines of background sources (e.g., \citealt{Wu93}), place this remnant at a distance of $2.18\pm0.08\rm~kpc$ \citep{Winkler03}, well consistent with the distance obtained from \ion{H}{1} observations \citep{Dubner02}. The apparent radio/X-ray diameter of SN1006 is $\sim30^\prime$, or $\sim19\rm~pc$ at this distance.

Similar as other young SNRs, the multi-wavelength properties of SN1006 also show clear spatial variations. \emph{First}, SN1006 can be apparently divided into two distinguishable parts. The northeast (NE) and southwest (SW) lobes are radio (e.g., \citealt{Reynolds86,Dyer09}), hard X-ray (e.g., \citealt{Koyama95,Winkler14}), and TeV (\citealt{Acero10}) bright, and the axis connecting the bright non-thermal regions is roughly aligned with the Galactic plane (\citealt{Gaensler98}). They are thought to be dominated by synchrotron emission of relativistic electrons accelerated at the SNR blast wave (e.g., \citealt{Koyama95}). On the other hand, most of the SNR interior, including the northwest (NW) shell, are dominated by thermal emission characterized by strong emission lines of heavy elements (e.g., \citealt{Koyama08,Uchida13}). \emph{Second}, not only the relative contribution of the two major components, but also the spectral properties of the non-thermal emission within the NE/SW lobes and the thermal emission within the SNR interior, show significant variations. These variations indicate the spatial variation of particle acceleration conditions (e.g., \citealt{Rothenflug04,CassamChenai08,Miceli13,Miceli14}), as well as the thermal and chemical properties across the entire remnant (e.g., \citealt{Uchida13,Winkler14}), which await better characterizations. \emph{Third}, the multi-wavelength properties of SN1006 also show noticeable spatial variations. In addition to the two prominent non-thermal lobes, the spatial variation of H$\alpha$ (e.g., \citealt{Winkler03,Raymond07,Nikolic13}), infra-red (IR; \citealt{Winkler13}), and \ion{H}{1} 21-cm line emissions (\citealt{Dubner02,Miceli14}) on both large (comparable to the size of the SNR) and small (comparable to the size of some prominent features) scales further indicates the considerable spatial variation of the density of the ambient interstellar medium (ISM), although SN1006 is believed to evolve in a relatively low-density and uniform environment.

We herein study SN1006 with our newly developed spatially resolved spectroscopy analysis method. We have obtained high quality X-ray data of SN1006 through our \emph{XMM-Newton} Large Program (LP) and some archival observations. The high sensitivity and large field of view (FOV) of \emph{XMM-Newton} help us to collect enough photons over the entire remnant, thus make SN1006 the best candidate to test how the spatially resolved spectroscopy analysis with our new techniques could improve our understanding of young SNRs. The paper is organized as follows: In \S\ref{PaperIsec:DataReduction}, we briefly introduce our \emph{XMM-Newton} LP and also the archival data used in this project. Basic data calibration is also detailed in this section. In \S\ref{PaperIsec:Methods}, we describe our new methods developed to conduct spatially resolved spectroscopy analysis. The parameter maps and other products of this analysis are presented and further discussed in \S\ref{PaperIsec:Results}. We summarize the main results and conclusions in \S\ref{PaperIsec:Summary}. More discussions on the thermal and non-thermal emissions of this remnant will be presented in companion papers.

\section{Observations and Data Calibration}\label{PaperIsec:DataReduction}

\subsection{XMM-Newton Large Program and Archival Data}\label{PaperIsubsec:Data}

In this work, we analyze the data obtained from the \emph{XMM-Newton} LP of SN1006 (PI: A. Decourchelle, $\sim700\rm~ks$ of total exposure time). These observations were all performed with the ``Medium'' filter by using the full-frame (FF) mode for the MOS (Metal Oxide Semi-conductor) cameras and the extended full-frame (EFF) mode for the PN cameras. We also select archival \emph{XMM-Newton} EPIC (the European Photon Imaging Camera) observations with pointing positions within $30^\prime$ from the center of SN1006. Only observations with at least one of the EPIC cameras operating in either FF or EFF mode are considered. All these observations are also carried out with the ``Medium'' filter. Information of all the selected observations is summarized in Table~\ref{PaperItable:observations}. Some analysis of these LP data have already been published in \citet{Miceli12,Miceli13,Miceli14,Broersen13}.

\subsection{Data calibration}\label{PaperIsubsec:Calibration}

We reduce the data based on \emph{XMM-Newton} Science Analysis Software (SAS) v12.0.1. For each observation listed in Table~\ref{PaperItable:observations}, the Observation Data Files (ODF) for each of the EPIC instruments (MOS-1, MOS-2 and PN) are reprocessed using the SAS tasks \emph{emchain} and \emph{epproc}. We identify and tag the low-energy noise in the MOS CCDs in anomalous states, i.e., with an elevated event rates between 0 and 1~keV, using the SAS task \emph{emtaglenoise}, following the algorithm described in \citet{Kuntz08}. All events in all CCDs tagged as noisy are then filtered out. CCDs in anomalous states with this low-energy noise are also summarized in Table~\ref{PaperItable:observations}. Each dataset is further screened for periods of soft proton flaring through the creation of full-field (but remove brightest point-like sources which may be time variable) light curves in broad band (0.3-12~keV for MOS and 0.3-14~keV for PN). Good Time Interval (GTI) selections are made by setting a threshold with $3~\sigma$ clipping for each instrument. The resulting effective exposure times for each instrument after this background flare filtering are listed in Table~\ref{PaperItable:observations}. The total effective exposure times of all the observations for MOS-1, MOS-2, and PN are 683, 710 and 439~ks, respectively. Out-of-Time (OoT) events have a non-negligible impact on the PN data and are also reprocessed with the SAS task \emph{epchain} and filtered in an identical way to the primary PN datasets. This OoT events are further subtracted in the following imaging and spectral analyses.

\begin{deluxetable}{lcccccccc}
\centering
\scriptsize 
  \tabletypesize{\scriptsize}
  \tablecaption{\emph{XMM-Newton} observations of SN1006 used in this work}
  \tablewidth{0pt}
  \tablehead{
 \colhead{ObsID} & \colhead{Start Date} & \colhead{noise CCD} & \colhead{$t_{liv,M1}$} & \colhead{$t_{liv,M2}$} & \colhead{$t_{liv,PN}$} & \colhead{$t_{eff,M1}$} & \colhead{$t_{eff,M2}$} & \colhead{$t_{eff,PN}$}
 }
\startdata
0077340101 & 2001-08-10 & -                      & 64862 & 64910 & 55980 & 30716 & 31148 & 20896 \\
0077340201 & 2001-08-10 & -                      & 57180 & 57213 & 44280 & 24678 & 24491 & 18969 \\
0111090101 & 2000-08-20 & -                      & 7425  & 7430  & 2954  & 7325  & 7430  & 2954  \\
0111090301 & 2000-08-17 & -                      & 5009  & 5062  & 899   & 1678  & 1679  & 0     \\
0111090601 & 2001-08-08 & -                      & 15770 & 15776 & 10387 & 7560  & 7957  & 4337  \\
0143980201 & 2003-08-14 & -                      & 30077 & 30089 & 23184 & 16896 & 16904 & 11915 \\
0202590101 & 2004-02-10 & -                      & 42796 & 42886 & 35155 & 29421 & 31563 & 19733 \\
0306660101 & 2005-08-21 & 4 of MOS-1, 5 of MOS-2 & 33424 & 33446 & 25526 & 9546  & 10528 & 6582  \\
0555630101 & 2008-08-22 & 5 of MOS-2             & 44851 & 44884 & 34437 & 44551 & 44781 & 26517 \\
0555630201 & 2008-08-04 & 4 of MOS-1             & 107848& 107885& 84401 & 94477 & 95999 & 52548 \\
0555630301 & 2009-02-19 & -                      & 119926& 119861& 93164 & 89856 & 91656 & 59983 \\
0555630401 & 2009-01-28 & 4 of MOS-1             & 102974& 103016& 81374 & 75749 & 81992 & 41791 \\
0555630501 & 2008-07-31 & 5 of MOS-2             & 125449& 125430& 100663& 85498 & 96372 & 51187 \\
0555631001 & 2008-08-28 & 5 of MOS-2             & 64641 & 64665 & 50191 & 60775 & 61571 & 44400 \\
0653860101 & 2010-08-28 & 4 and 5 of MOS-1       & 124863& 124952& 101182& 104396& 105661& 77326 \\
Total & - & - & 947095 & 947505 & 743777 & 683122 & 709731 & 439140
\enddata
\tablecomments{\scriptsize The third column summarizes the identification number of MOS CCD(s) with low-energy noise. $t_{liv,M1}$, $t_{liv,M2}$, and $t_{liv,PN}$ are the dead-time corrected on time (LIVETIME) for EPIC MOS-1, MOS-2, and PN, respectively. $t_{eff,M1}$, $t_{eff,M2}$, and $t_{eff,PN}$ are the effective exposure times of EPIC MOS-1, MOS-2, and PN after background flare removal. The last row is the total exposure time of individual cameras of all the data used in this work. All the exposure times are in unit of second.}\label{PaperItable:observations}
\end{deluxetable}

The instrument (telescope+detector) response is not flat, i.e., the effective area at a given energy depends on the position in the focal plane. This vignetting effect is extremely important for the analysis of inhomogeneous extended source such as SN1006. We apply the SAS task \emph{evigweight} to both the OoT and the primary events lists, weighting each EPIC events with inverse effective area over one exposure, so that the derived event list is equivalent to what one would get for a flat instrument. This allows the use of single on-axis Ancillary Response Files (ARF) for each instrument, and does not require further effective area corrections in the following imaging and spectral analyses.

\begin{figure}[!h]
\begin{center}
\epsfig{figure=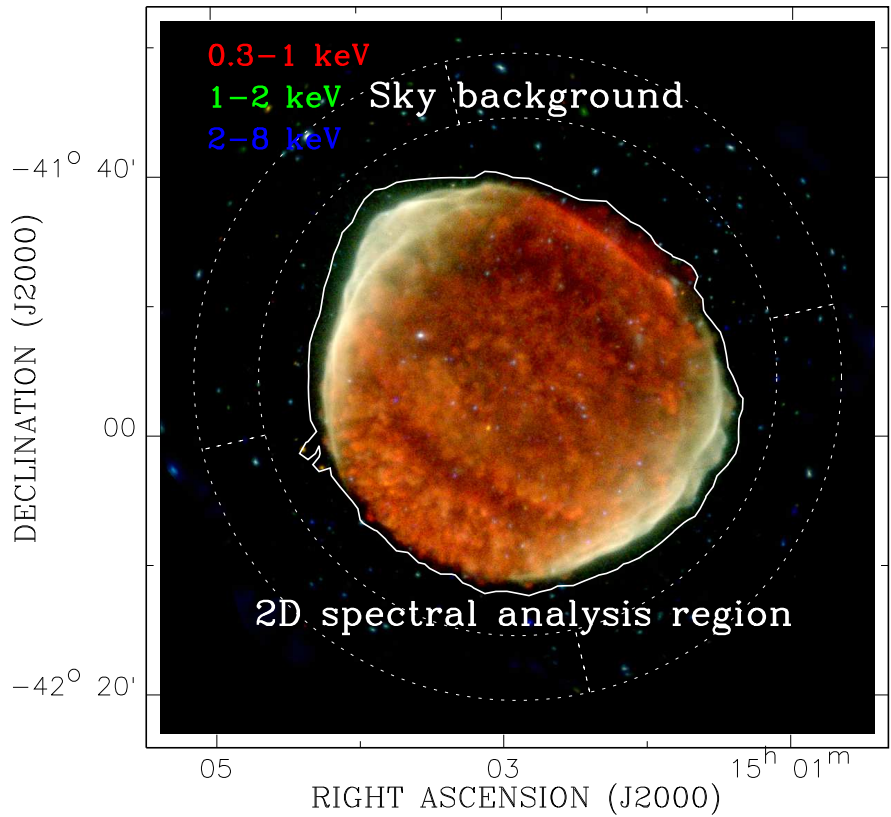,width=0.7\textwidth,angle=0, clip=}
\caption{Broad band tricolor images (Red: 0.3-1~keV; Green: 1-2~keV; Blue: 2-8~keV) of the entire field of view of the \emph{XMM-Newton} observations of SN1006. The white solid contour shows the region to extract spectra for the whole remnant and to do spatially resolved spectroscopy analysis. The white dashed annulus show the region to extract the sky background spectra, and is divided into four quadrants to study the possible azimuthal variation of the sky background (Appendix~\ref{PaperIAppendix:SkyBackground}). To clarify, removed point-like sources are not shown.}\label{PaperIfig:tricolorregions}
\end{center}
\end{figure}

The source event maps and the instrument exposure maps from each observation and each camera are constructed in several bands. In order to correct for the OoT events in PN observations, we also construct OoT images and scale them to the expected OoT event fraction (6.3\% in FF and 2.2\% in EFF modes respectively, according to \emph{XMM-Newton} Users Handbook: \url{http://xmm.esac.esa.int/external/xmm_user_support/documentation/uhb/index.html}). These OoT images are then subtracted from the raw PN images. A subtraction of the instrumental background is applied to images from all instruments by making use of the EPIC Filter Wheel Closed (FWC) data obtained from the \emph{XMM-Newton} background analysis website (\url{http://xmm2.esac.esa.int/external/xmm\_sw\_cal/background/filter\_closed/index.shtml}). These FWC data are reprocessed and renormalized to match the 10-12/12-14~keV (for MOS/PN) counts number of the source images before subtraction.

The source, exposure, and background maps from different observations are then combined together using the SAS task \emph{emosaic}. All the images are binned to a pixel size of $3.2^{\prime\prime}$, slightly smaller than the FWHM (Full Width at Half Maximum) of the \emph{XMM-Newton} mirror on-axis Point Spread Function (PSF; $\sim6^{\prime\prime}$). Therefore, we do not lose resolution in constructing images. The background-subtracted event map is further adaptively smoothed with the SAS task \emph{asmooth} to a desired signal-to-noise ratio of 5. The exposure map is smoothed according to the same template as the event map. We then produce final background-subtracted and exposure-corrected flux images with these smoothed images (e.g., Fig.~\ref{PaperIfig:tricolorregions}).

Because there exist a lot of soft X-ray knots in SN1006, which may be mis-identified as point sources in usual source detection tools, we conduct point-source detection only in the hard X-ray band (2-8~keV). We create a constant PSF map of $10^{\prime\prime}$ for the mosaiced images, which is good enough for a simple detection and removal of them for the study of diffuse emission. We then adopt this PSF map to a standard wave detection tool \emph{wavdetect}. We finally inspect the detected point-like sources visually to remove obvious false detections. The point sources are only removed in extracting the background spectra (Appendix~\ref{PaperIAppendix:SkyBackground}). There are many faint point-like sources projected inside the SNR. The X-ray emission of most of these sources peak at hard X-ray so do not contribute significantly at $\lesssim2\rm~keV$ which we most concern in this paper. Therefore, we do not remove these point sources from the source spectra. We caution that some peculiar features on the parameter maps shown in the following sections may be caused by foreground or background point sources instead of the emission from the SNR itself. This contamination is in general not significant, but could be important when the truly diffuse emission is faint (e.g., close to the two point sources at ${\rm RA, Dec}\thickapprox 15^h04^m20^s, -42^\circ02^\prime$ outside the southeast rim of the SNR; Fig.~\ref{PaperIfig:2DSpec_meshes}).

\section{Spatially-Resolved Spectroscopy}\label{PaperIsec:Methods}

\subsection{Mapping the spectral parameters: general procedure}\label{PaperIsubsec:2DSpecProcedure}

X-ray observations typically provide us with multi-dimensional data, i.e., by recording the (x,y) position on the focal plane, the arrival time, and the energy of the photons. This multi-dimensional data allows us to study the spatial distribution of many physical parameters, in addition to simply constructing broad- or narrow-band intensity images. Spatially-resolved spectroscopy represent such techniques to map out the physical parameters of X-ray bright extended sources. Different from previous studies which often extract spectra from several interesting regions of an extended source, we herein introduce a new method to directly map out the spectral analysis parameters in many tessellated regions. Similar techniques have been discussed in several literatures. For example, \citet{Randall08} used two different methods (the over-sampling of the image at each pixel or the tessellated mesh) to map out the temperature distribution around the Virgo cluster galaxy M86.

\subsubsection{Creating tessellated meshes}\label{PaperIsubsubsec:Mesh}

\begin{figure}[!h]
\begin{center}
\epsfig{figure=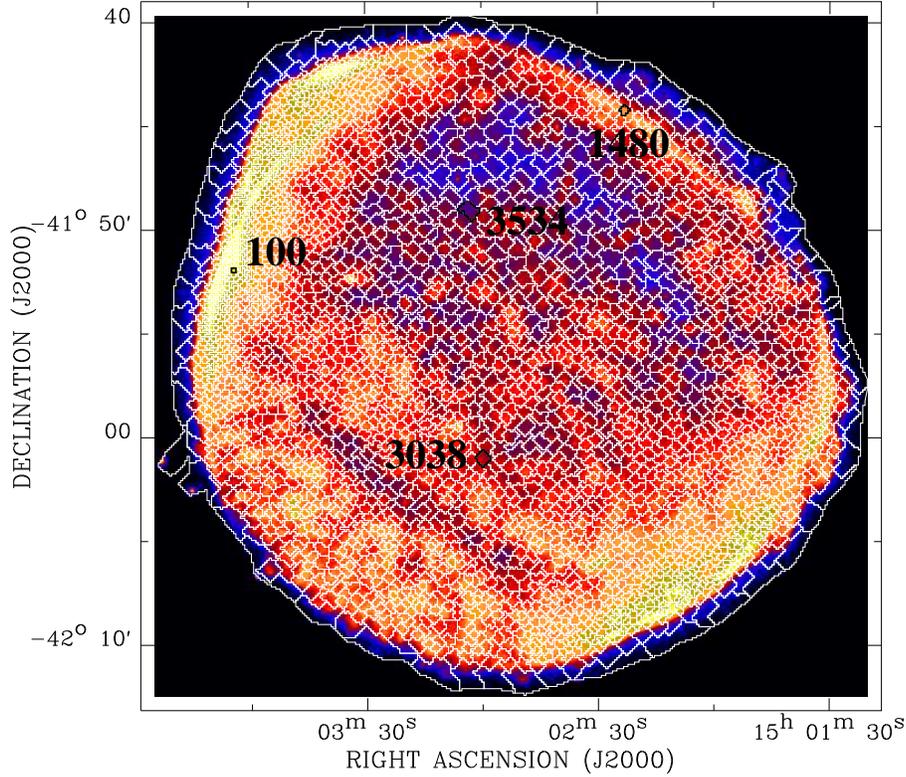,width=0.7\textwidth,angle=0, clip=}
\caption{Tessellated meshes used for spatially resolved spectroscopy analysis (the high resolution run) overlaid on the 0.3-8~keV image of SN1006. There are in total 3596 regions, each containing $\gtrsim 10^4\rm~counts$ from the combination of MOS-1, MOS-2, and PN. The four black meshes are the regions used to extract the sample spectra in Fig.~\ref{PaperIfig:spec_mesh}, with the region numbers denoted beside.}\label{PaperIfig:2DSpec_meshes}
\end{center}
\end{figure}

We adopt a new algorithm in order to construct tessellated meshes adaptively. We first find the brightest unbinned pixel in the original point-source-removed broad-band (0.3-8~keV) counts image and put this pixel into a new mesh. If the pixels already included in the mesh contain total (MOS-1+MOS-2+PN) counts number above our threshold of $10^4\rm~counts$, we create one polygon region containing all the pixels in the mesh. If the pixels contain less counts, we then add the brightest pixel surrounding the previous added pixel to the mesh and compare the total counts to the threshold again. This process is repeated until the mesh contains more counts than the threshold or no surrounding pixel is unbinned. Isolated pixels between different meshes are leaving unbinned during the creation of the meshes. These unbinned pixels are finally individually merged into the neighboring meshes with the smallest average distance to them. This algorithm creates tessellated polygon regions each has at least a counts number of the input threshold, while each pixel only belongs to one single region (the meshes). An example of the created meshes (3596 tessellated regions), as will be adopted in the following analysis, is shown in Fig.~\ref{PaperIfig:2DSpec_meshes}. The smallest meshes have typical diameter of 4~pixel, or $\sim13^{\prime\prime}$, larger than the FWHM of the PSF. Therefore, the tail of the PSF should have little effect on the spectral analysis.

\subsubsection{Extracting spectra of individual regions}\label{PaperIsubsubsec:SpecExtraction}

We extract spectra of MOS-1, MOS-2, and PN from each of the polygon regions of each observations containing positive counts numbers. We then weight these spectra as well as the corresponding background and response files and stack them together for MOS-1, MOS-2, and PN, respectively. Typically, the spectrum of each instrument from a single region contains several thousand net counts, enough for the purpose of roughly characterizing the soft X-ray spectral energy distribution. To save computer time and space, we do not create response files [the Redistribution Matrix File (RMF) and the ARF] for all the spectra extracted for the three instruments of the 15 observations from the 3596 regions ($\sim10^5$ spectra and the corresponding RMF; hereafter referred as the high resolution run). Instead, we create lower resolution meshes (with 182 regions in total, hereafter referred as the low resolution run), thus with higher counting statistic, and create RMFs for the spectra extracted from each regions of the three instruments of all the observations. The RMFs are then weighted according to the effective exposure time of individual spectra and merged together to create single weighted RMF for each region. These RMFs are taken as templates in the high resolution run. In the high resolution run, for each region, we find the nearest region from it in the low resolution run and directly use the existing RMF from this region. Because we have already corrected the vignetting in the calibration (\S\ref{PaperIsubsec:Calibration}), we use identical on-axis ARF of each instrument for all the regions. In both low and high resolution runs, we still extract the source and background spectra separately for individual regions, instruments, and observations, and stack them according to their effective exposure time and area scale, similar as the standard way typically taken by many other authors. The use of template RMF may cause some biases in the spectral analysis, but we have double checked all the prominent features in the spectra extracted from the tessellated meshes, by examining the spectra extracted from larger regions with the response files individually generated (\S\ref{PaperIsubsec:ReliabilityParaMap}).

\subsection{Spectral modeling of individual regions}\label{PaperIsubsec:SpecModelMesh}

\subsubsection{Spectral model}\label{PaperIsubsubsec:SpecModel}

Analysis of the sky background is presented in Appendix~\ref{PaperIAppendix:SkyBackground}. We add the sky background components (with all the parameters fixed) to the model of the source spectra. In order to do this, the normalization of the three sky background components (VMEKAL, power law, and MEKAL) are rescaled according to the ratio of the effective areas of the source and sky background regions. In the analysis of both source and sky background spectra, we adopt the FWC data as the instrumental background and have subtracted them before spectral analysis.

We fit the source spectra with a ``VNEI+SRCUT'' model. The VNEI component (Non-Equilibrium Ionization collisional plasma model, with parameters describing the relative abundances between different elements) describes the thermal plasma contribution, assuming a constant temperature and single ionization parameter. The SRCUT component describes the synchrotron emission from an exponentially cutoff power-law distribution of electrons in a homogeneous magnetic field (\citealt{Reynolds99}). 

Both the VNEI and SRCUT components are subjected to foreground absorption described with a WABS model (photo-electric absorption using Wisconsin cross-sections). 
It is possible that the distribution of the ambient ISM is inhomogenous, which could affect the absorption and thus the shape of the soft X-ray spectra. However, based on \citet{Dubner02}'s \ion{H}{1} column density map in the surrounding region of SN1006, the variation of absorbing \ion{H}{1} column density ($N_{\rm H}$) is at most $\sim25\%$ (also see \citealt{Miceli14}). Because $N_{\rm H}$ in the direction of SN1006 is quite low, such small variations on $N_{\rm H}$ does not significantly affect the spectral fitting results. We therefore fix the absorption column density at the foreground value of $N_{\rm H}=6.8\times10^{20}\rm~cm^{-2}$ \citep{Dubner02}.

\begin{figure}[!h]
\begin{center}
\epsfig{figure=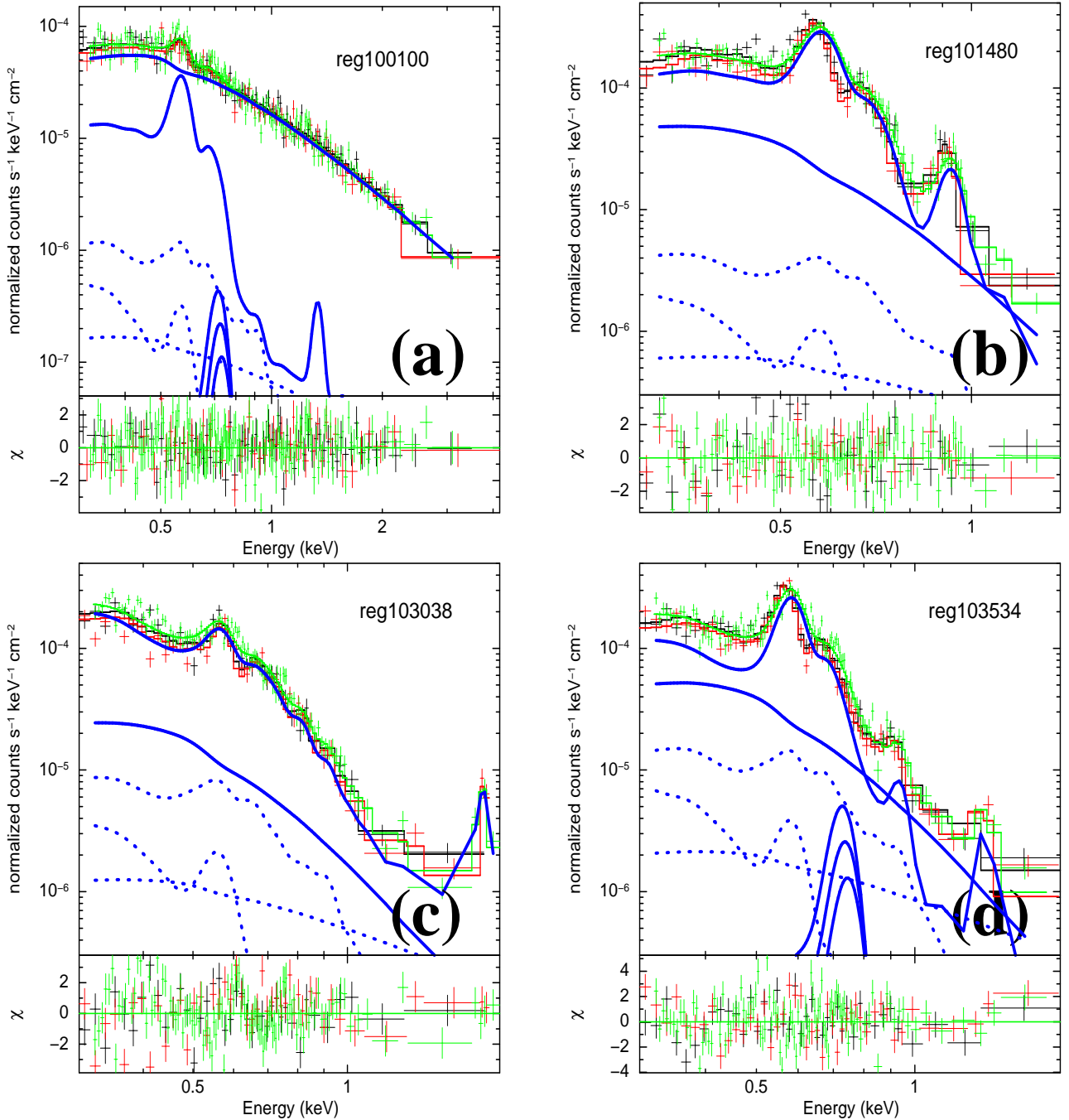,width=1.0\textwidth,angle=0, clip=}
\caption{Examples of spectral fitting of individual tessellated regions (region numbers denoted in Fig.~\ref{PaperIfig:2DSpec_meshes}). The black, red, and green data points are the instrumental background-subtracted spectra of MOS-1, MOS-2, and PN, while the solid curves with the corresponding colors are their best-fit models. The vertical axis of the upper half of each panels is the intensity \emph{per unit effective area} (different from other spectra shown in this paper), so the spectra of different instruments have roughly identical intensities. The blue solid curves are the source components (VNEI, SRCUT, and Gaussian lines) of the PN spectrum, while the dashed curves are the sky background components (the local hot bubble, Galactic halo, and distant AGNs; Appendix~\ref{PaperIAppendix:SkyBackground}). The lower panels show the residuals in terms of $\sigma$. Panel~(a) is dominated by non-thermal emission, while panels~(b-d) are dominated by thermal emission, with strong Ne, Si, and O lines, respectively. Panel~(d) also has a relatively large contribution from the \ion{O}{7}~K$\delta,\epsilon,\zeta$ lines (the three Gaussian lines marked in blue solid curves), while this component is too weak to appear on the plot in panels~(b) and (c).}\label{PaperIfig:spec_mesh}
\end{center}
\end{figure}

For the VNEI component, we adopt the Atomic Data for Astrophysicists (AtomDB; \citealt{Foster12}). We caution that the VNEI model only has a single ionization parameter ($n_et$), so does not take account the possibly different ionization age of different elements or a distribution of ionization age in the post-shock region. The ionization age distribution of the post-shock plasma, and/or the different temperatures of the ions and electrons, are better described with a plane-parallel shock plasma model (VPSHOCK) or an X-ray spectral model describing SNRs in the Sedov-Taylor phase (Sedov model; \citealt{Borkowski01}). However, spectral fitting with such models take too long computer time for the 3596 regions, so are not adopted here. 

Our single temperature (1-T) VNEI model may also fail to describe the possible multi-temperature structure of the plasma (e.g., \citealt{Uchida13}). However, most of the spectra from individual meshes have poor counting statistic at high energy (e.g., at the Si line band), so have poor constraint on the properties of some possible high-temperature components. Therefore, the spectral model adopted in the present paper is only aimed at roughly decomposing the thermal and non-thermal contributions, as well as characterizing the average thermal and ionization states of the plasma. We will briefly discuss the thermal structure of the plasma in \S\ref{PaperIsubsubsec:2DSpec_kT}. Further discussions on modeling the thermal spectra of SN1006 based on higher-quality spectra but lower-resolution meshes will be presented in companion papers. 

We set the O, Ne, Mg, Si, and Fe abundances free, the S abundance equal to Si, and Ni abundance equal to Fe. All these elements produce strong emission lines in soft X-ray, but Fe and Ni do not have well separated emission line bumps at $\lesssim2\rm~keV$, in which our spectra typically have enough counts (e.g., Fig.~\ref{PaperIfig:spec_mesh}).

We further add into our models three narrow gaussian lines (line width fixed at $10^{-4}\rm~keV$) at 0.714~keV, 0.723~keV, and 0.730~keV to account for the higher level transitions (K$\delta$, K$\epsilon$, K$\zeta$) of \ion{O}{7} lines which are not included in the VNEI code (e.g., Fig.~\ref{PaperIfig:spec_mesh}a,d). Following \citet{Yamaguchi08}, we assume K$\epsilon$/K$\delta$=K$\zeta$/K$\epsilon$=0.5, so there is only one free parameter of this component (the normalization of the \ion{O}{7} K$\delta$ line). At temperature $kT\gtrsim0.6\rm~keV$, the \ion{O}{7} lines are typically weak compared to the \ion{O}{8} lines. At lower temperatures where the \ion{O}{7} line is dominant, the line flux decreases rapidly along the K-shell transition series, so in most cases only K$\alpha-\gamma$ transitions are important. However, SN1006 has a low ionization age and fairly high temperature (\S\ref{PaperIsec:Results}). It is thus possible that these high level transitions are fairly strong in some regions, the spectra of which show clear residual at 0.7-0.75~keV without adding these gaussian lines. Alternatively, the spectral fitting residual at 0.7-0.75~keV could also be explained as the Fe~L shell transitions from low ionization ions (less than Fe$^{16+}$; e.g., \citealt{Warren04,Uchida13}). We will further discuss this possibility in \S\ref{PaperIsubsubsec:metal_FeK}. Nevertheless, for the convenience of presentation, we will call these added gaussian lines the \ion{O}{7}~$\rm K\delta-\zeta$ lines throughout the paper, regardless of its real origin from oxygen or iron.

In the SRCUT model, we fit the radio-to-X-ray photon index $\alpha$ and the cutoff frequency $\nu_{cutoff}$, but fix the normalization using the radio flux obtained from \citet{Dyer09}. We assume a constant radio spectral slope of 0.5 ($\alpha\sim0.45-0.6$; \citealt{Dyer01,Acero07,Katsuda10}), in order to convert to the flux at 1.4~GHz in \citet{Dyer09} to the flux at 1~GHz adopted in the SRCUT model. Small variation of the assumed $\alpha$ does not significantly affect our spectral analysis results.

\begin{figure}[!h]
\begin{center}
\epsfig{figure=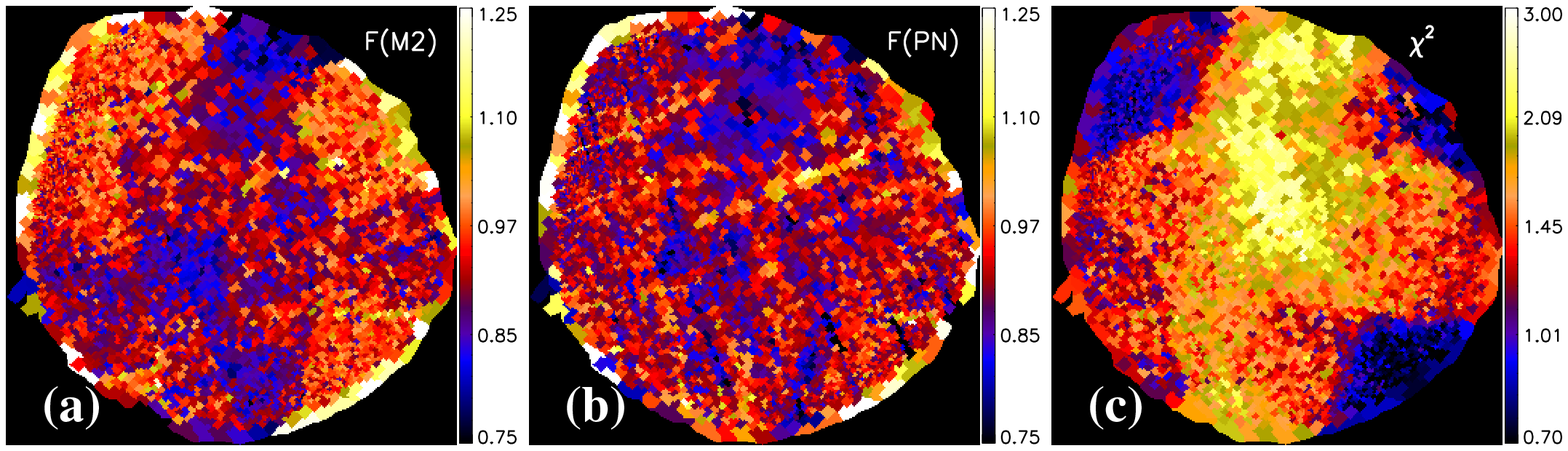,width=1.0\textwidth,angle=0, clip=}
\caption{Panels (a) and (b): Normalization factor of MOS-2 (a) and PN (b) to the MOS-1 spectra, in order to account for the possible difference between the spectra extracted from different instruments. Panel (c): $\chi^2/\rm d.o.f$ ($\rm d.o.f$ is the degree of freedom) of the spectral fitting of each tessellated regions.}\label{PaperIfig:fitgoodness}
\end{center}
\end{figure}

The spectra of the three EPIC camera (MOS-1, MOS-2, and PN) are jointly fitted with all the parameters linked, except for a constant normalization factor subjected to all the model components. This normalization factor is used to account for the possible difference in area scale and calibration bias of the spectra extracted from different instruments. As shown in Fig.~\ref{PaperIfig:fitgoodness}a,b, it is close to 1 over the entire remnant, with slightly larger bias appears near the gaps between different CCD chips. This bias will not significantly affect our results, however. Therefore, in the following analyses and discussions of this paper, we will directly use the parameters from jointly fitting the MOS-1, MOS-2, and PN spectra, without considering the normalization factor. 

We also add a gain correction to the response file of the PN spectrum (adopting the ``gain'' model in XSpec). This is aiming at accounting for the deficiency in the low energy calibration of the PN camera (\citealt{Dennerl04}). The slope of ``gain'' is fixed at 1 and the offset is set free.

Such a ``VNEI+SRCUT+\ion{O}{7}~K$\delta,\epsilon,\zeta$~line+Background'' model typically gives acceptable spectral fitting results. Examples of spectra extracted from individual regions dominated by different spectral features are shown in Fig.~\ref{PaperIfig:spec_mesh} and the best-fit spectral parameters of them are summarized in Table~\ref{PaperItable:ExampleSpecPara} together with their statistical errors. The maximum $\chi^2/{\rm d.o.f}$ for individual tessellated meshes is $\sim2.81$, and $\gtrsim93\%$ of the regions have $\chi^2/{\rm d.o.f}<2.0$ (Fig.~\ref{PaperIfig:fitgoodness}c). We have individually inspected all the spectra with $\chi^2/{\rm d.o.f}\geq2.0$, and refit the spectra if necessary. In most of the cases, the large $\chi^2/{\rm d.o.f}$ value could be explained by the too simple 1-T spectral model adopted in the present paper.

\begin{deluxetable}{lccccccccccc}
\centering
\scriptsize 
  \tabletypesize{\scriptsize}
  \tablecaption{Examples of spectral fitting parameters and errors}
  \tablewidth{0pt}
  \tablehead{
  \colhead{Region} & \colhead{kT} & \colhead{O} & \colhead{Ne} & \colhead{Mg} & \colhead{Si} & \colhead{Fe} & \colhead{$\log n_et$} & \colhead{$norm_{\rm VNEI}$} & \colhead{$\Gamma$} & \colhead{$\nu_{cutoff}$} \\ 
 & keV & solar & solar & solar & solar & solar & $\log (\rm cm^{-3}s)$ & & & $10^{6}\rm~GHz$
 }
\startdata
100        &  $>0.22$ & $0.84_{-0.27}^{+1.17}$ & $<0.26$ & $<34.4$ & $<28.0$ & $<13.5$ & $9.49_{-0.33}^{+1.22}$ & $1.00_{-0.49}^{+0.25}$ & $0.49\pm0.01$ & $46.4_{-11.3}^{+7.6}$ \\
1480      &  $0.23_{-0.04}^{+0.09}$ & $1.10_{-0.20}^{+0.36}$ & $1.95_{-0.58}^{+0.82}$ & $<7.43$ & $<3.16$ & $<0.17$ & $>10.07$ & $34.2_{-5.1}^{+6.6}$ & $<0.15$ & $0.91_{-0.07}^{+0.76}$ \\
3038      & $1.47_{-0.22}^{+0.33}$ & $0.59_{-0.26}^{+0.49}$ & $0.28\pm0.06$ & $0.59_{-0.44}^{+0.45}$ & $25.1_{-8.0}^{+25.4}$ & $1.57_{-0.51}^{+2.15}$ & $9.57_{-0.08}^{+0.05}$ & $4.41_{-2.07}^{+1.69}$ & $<0.14$ & $0.73_{-0.08}^{+0.16}$ \\
3534      & $1.30_{-0.23}^{+0.73}$ & $<19.5$ & $6.36_{-1.47}^{+1.59}$ & $50.7_{-26.2}^{+16.2}$ & $420_{-260}^{+98}$ & $<6.96$ & $9.43_{-0.04}^{+0.06}$ & $0.26_{-0.02}^{+0.04}$ & $0.116_{-0.004}^{+0.032}$ & $0.85_{-0.04}^{+0.21}$
\enddata
\tablecomments{\scriptsize Some key spectral fitting parameters and their 90\% confidence errors of the spectra shown in Fig.~\ref{PaperIfig:spec_mesh}. $norm_{\rm VNEI}$ is in unit of  $\frac{10^{-20}}{4\pi[D_A(1+z)]^2}\int n_en_{\rm H}dV$. See the text in \S\ref{PaperIsubsubsec:DerivedPara} for details.}\label{PaperItable:ExampleSpecPara}
\end{deluxetable}

\subsubsection{Derived parameters}\label{PaperIsubsubsec:DerivedPara}

We further derive some physical parameters based on the parameters directly obtained from spectral fitting. We first estimate the electron number density ($n_e$) from the emission measure of the plasma. In the VNEI model, the normalization is defined as: $norm=\frac{10^{-14}}{4\pi[D_A(1+z)]^2}\int n_en_{\rm H}dV$, where $D_A$ is the distance to the source in cm, $n_{\rm H}$ is the hydrogen number density, and $dV$ is the volume of the projected spectral analysis regions. Adopting an outer shell radius of 9.5~pc (\S\ref{PaperIsec:Introduction}) and an ambient medium density of $n_0\sim0.05\rm~cm^{-3}$ (e.g., \citealt{Miceli12}) and solar abundance, the total swept up ISM mass by SN1006 will be $\sim5\rm~M_\odot$, not much higher than the total ejecta mass of a Type~Ia SN. Therefore, we may expect a significant metal enrichment by SN ejecta, as indicated by the significant spatial variations of different elements (see \S\ref{PaperIsec:Results}). However, in the present paper, we do not decompose the ISM and ejecta components in our simple 1-T thermal plasma model. In most of the spectra dominated by thermal emission, O lines are the most prominent (e.g., Fig.~\ref{PaperIfig:spec_mesh}), and the fitted O abundances are close to solar (\S\ref{PaperIsubsubsec:metal_Oline}). Therefore, for the mixed ISM plus ejecta spectra, we assume solar abundance in the calculation of $n_e$ for simplicity. The assumption of solar abundance links the electron and hydrogen densities by $n_e=1.21n_{\rm H}$. 

\begin{figure}[!h]
\begin{center}
\epsfig{figure=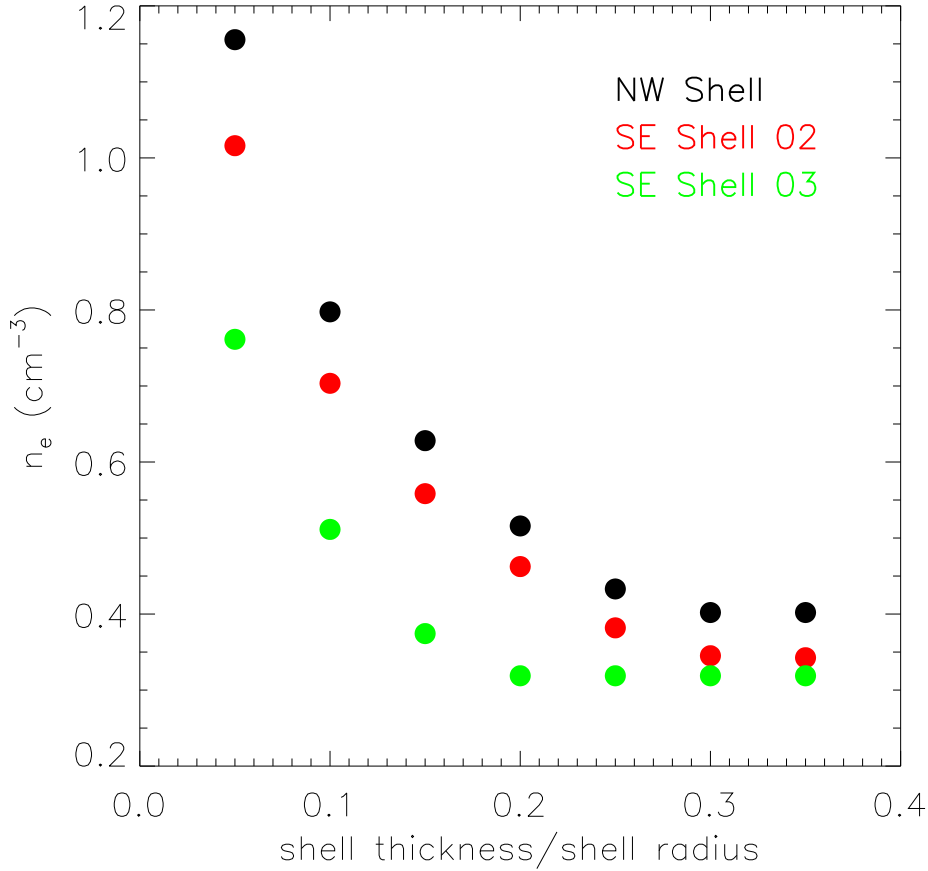,width=0.7\textwidth,angle=0, clip=}
\caption{Derived post-shock electron number density in several outer regions v.s. the assumed thickness of the thermal X-ray emitting shell (in unit of the outer radius of the shell). Region names are denoted in Fig.~\ref{PaperIfig:regionsindividual}.}\label{PaperIfig:ShellThickness}
\end{center}
\end{figure}

We adopt the spherical shell geometric model described in \citet{Miceli12} to estimate the volume of the spectral regions. We assume SN1006 has an outer radius of 11~pc, slightly larger than the average X-ray radius of $\sim9.5\rm~pc$ (\S\ref{PaperIsec:Introduction}), in order to cover all the X-ray emitting regions (the two non-thermal ``ears'' are slightly more extended). We also equalize each tessellated mesh to a sector-shaped region with the same area for the convenience of calculating its volume. There are two poorly constrained parameters of this model: the thickness of the thermal X-ray emitting shell and the volume filling factor $f$ within the X-ray emitting regions. We assume different shell thickness and estimate the plasma density accordingly. In Fig.~\ref{PaperIfig:ShellThickness}, we plot $n_e$ of some outermost regions dominated by thermal emission against the assumed shell thickness. 

The choice of regions dominated by thermal emission reduces the uncertainty caused by the highly variable shock compression ratio in regions with strong hadronic particle acceleration (e.g., \citealt{Decourchelle00,Miceli09,Vink10,Ferrand12,Ressler14}). We could therefore convert the measured post-shock plasma density to the pre-shock ambient ISM density by assuming a compression ratio of 4. This allows us to compare our results to the ambient ISM density inferred from multi-wavelength observations (e.g., \citealt{Dubner02,Raymond07,Heng07,Winkler13}). We caution that in our simple spectral analysis with the 1-T plasma model, we are not able to decompose the ISM and ejecta components, which could result in an overestimate of the ISM density. For example, in similar X-ray measurements of the SE shell, the typical post-shock electron density is $n_e\sim0.2\rm~cm^{-3}$, corresponding to a pre-shock ISM density of $n_0\sim0.05\rm~cm^{-3}$ (e.g., \citealt{Acero07,Miceli12,Winkler14}). We hereafter choose a shell thickness of 0.2 shell radius, which also produces a density of the NW shell roughly consistent with previous observations ($n_e\lesssim1\rm~cm^{-3}$).

As we have adopted a simple 1-T model, we further assume the plasma is volume filling within the X-ray emitting regions, i.e., $f\sim1$. If the plasma is highly structured below the resolving power of our tessellated meshes, $f$ will be significantly $<1$. In this case, the estimated $n_e$ will be lower limit of the real value.

From the electron density and the fitted ionization parameter $n_et$, we could further derive the ionization age $t_{ion}=n_et/n_e$. In this paper, we call $n_et$ the ionization parameter, different from some literatures where $n_et$ is called the ionization age, in distinguish with $t_{ion}$ with truly time dimension (in unit of year). As will be discussed in \S\ref{PaperIsubsubsec:2DSpec_kT}, the maximum value of $t_{ion}$ in the SNR interior is typically $\gtrsim500\rm~yr$, consistent with the age of SN1006 ($\sim10^3\rm~yr$ based on historical records; \citealt{Stephenson10}). This also indicates that the above geometric model and our estimates of $n_e$ are generally reliable.

It has been proved that the thermal X-ray emission is ejecta-dominated in some regions (e.g., the SE shell) while ISM-dominated in other parts of the SNR (e.g., the NW shell; \citealt{Acero07,Miceli09,Miceli12,Nikolic13,Winkler14}). The mixture of ISM and ejecta will not only affect the assumed abundance as discussed above, but also result in different shell thickness in different regions. Furthermore, the environmental density, as well as the filling factor of ISM and ejecta are definitely not uniform across the SNR. Therefore, we caution that the above estimates of $n_e$ and $t_{ion}$ based on the assumption of ISM abundance, uniform shell-like geometric model, and volume-filling thermal plasma can be largely biased in some specific regions. 

\subsection{Equivalent width map}\label{PaperIsubsec:2DSpecEW}

Equivalent width maps represent images of emission-line-to-continuum ratio, which could account for the locally variable underlying continuum and further reveal the distribution of truly line strength throughout extended X-ray sources over a wide range of surface brightness (e.g., \citealt{Hwang00}).

\subsubsection{Prominent emission lines and the corresponding energy ranges}\label{PaperIsubsubsec:Narrowbands}

\begin{figure}[!h]
\begin{center}
\epsfig{figure=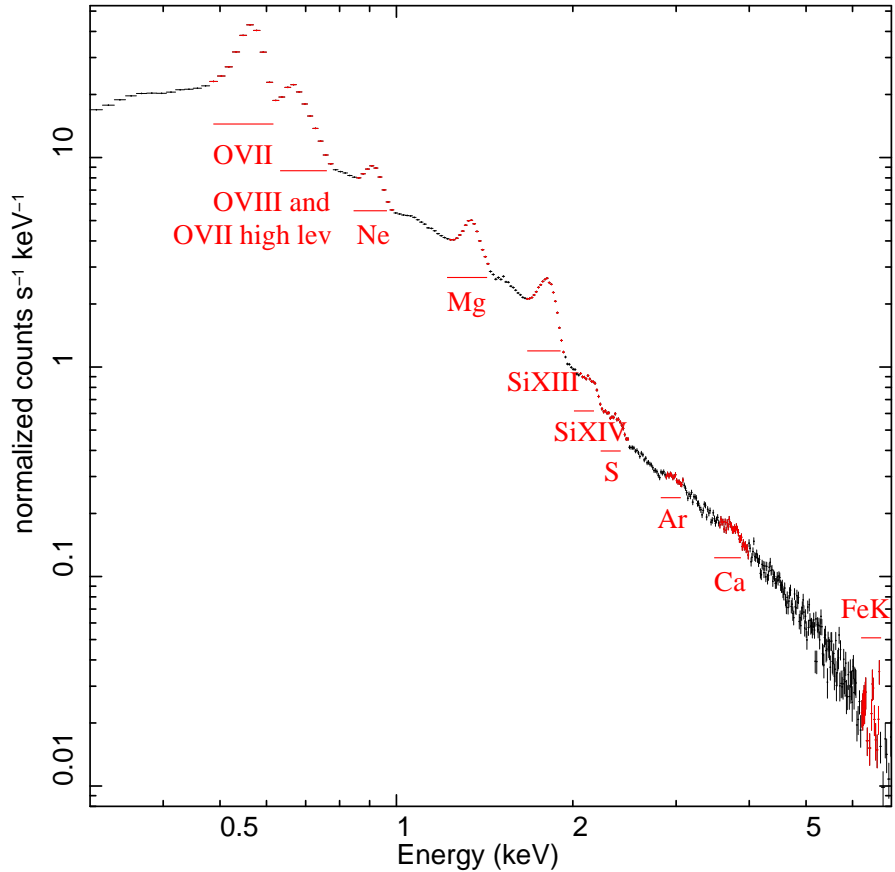,width=0.7\textwidth,angle=0, clip=}
\caption{MOS-1 spectrum extracted from the whole remnant (the white solid contour in Fig.~\ref{PaperIfig:tricolorregions}). The energy range and name of the emission lines are marked in red, with detailed information summarized in Table~\ref{PaperItable:EWmaps}. ``\ion{O}{7} high lev'' means the \ion{O}{7}~$\rm K\delta-\zeta$ lines which are blended with the \ion{O}{8} lines in this figure.}\label{PaperIfig:EWband}
\end{center}
\end{figure}

In order to define the energy ranges dominated by emission lines and continuum, we present the MOS-1 spectrum extracted from the entire remnant in Fig.~\ref{PaperIfig:EWband} (enclosed by the white solid contour in Fig.~\ref{PaperIfig:tricolorregions}). This spectrum features prominent emission line bumps of O, Ne, Mg, and Si, as well as weak features of S, Ar, Ca, and the K-shell blend of Fe. The \ion{O}{7}~$\rm K\delta-\zeta$ lines are blended with the \ion{O}{8} lines in Fig.~\ref{PaperIfig:EWband}. The Fe~L lines often do not have a prominent bump. The Fe~K lines are unambiguously detected and better resolved by \citet{Yamaguchi08,Uchida13} with the \emph{Suzaku} observations. We define the energy ranges of the line and continuum visually and mark the emission lines in red in Fig.~\ref{PaperIfig:EWband}. The energy ranges used to construct EW maps are summarized in Table~\ref{PaperItable:EWmaps}. We do not include the weak Ar and Ca lines in either the line or continuum bands.

\begin{deluxetable}{lccccccccc}
\centering
\scriptsize 
  \tabletypesize{\scriptsize}
  \tablecaption{Narrow bands used to created EW maps}
  \tablewidth{0pt}
  \tablehead{
  \colhead{} & \colhead{\ion{O}{7}} & \colhead{\ion{O}{8}} & \colhead{\ion{O}{7}~$\rm K\delta-\zeta$} & \colhead{Ne} & \colhead{Mg} & \colhead{\ion{Si}{13}} & \colhead{\ion{Si}{14}} & \colhead{\ion{S}{15}} & \colhead{Fe~K}
 }
\startdata
Line        & 470-650 & 650-700 & 700-780 & 850-1000  & 1230-1450 & 1660-1950 & 2050-2250 & 2250-2500 & 6200-6700 \\
$C_{low}$   & 300-470 & 300-470 & 300-470 & 780-850   & 1000-1230 & 1450-1660 & 1950-2050 & 1950-2050 & 4000-6200 \\
$C_{high}$  & 780-850 & 780-850 & 780-850 & 1000-1230 & 1450-1660 & 1950-2050 & 2500-2870 & 2500-2870 & 6700-7000
\enddata
\tablecomments{\scriptsize Line, $C_{low}$, and $C_{high}$ are the energy ranges in eV for the line, the low and high energy continuum to construct the EW maps, respectively. \ion{O}{7}~$\rm K\delta-\zeta$ is the \ion{O}{7}~K$\delta$, K$\epsilon$, and K$\zeta$ lines which are not included in the VNEI code. These lines are often not clearly separated from the \ion{O}{8} lines in the spectra of individual tessellated regions.}\label{PaperItable:EWmaps}
\end{deluxetable}

\subsubsection{Limitations of the abundance map and the continuum estimation from linear interpolation}\label{PaperIsubsubsec:LinearInterp}

In the simplest case, we obtain the continuum under the emission lines with a linear interpolation between the low and high energy continuums just beside the lines ($C_{low}$ and $C_{high}$ in Table~\ref{PaperItable:EWmaps}). The EW is calculated by subtracting this linearly interpolated continuum from the flux in the line band (Table~\ref{PaperItable:EWmaps}) and then divide the same continuum.

Both the abundance maps created with the spatially-resolved spectroscopy method (\S\ref{PaperIsubsec:2DSpecProcedure}) and the EW maps constructed with this linear interpolation method can reveal the spatial distribution of metal-enriched gas. However, both methods have their own disadvantages. The accuracy of the \emph{abundance map} is closely related to the goodness of spectral fitting. In many cases there are significant residuals at some emission lines, especially when the counting statistic at these lines are poor (e.g., in many cases, the Si lines are not resolved in the spectra of individual meshes). In addition, the tessellated meshes, constructed for spectral analysis, also have lower angular resolution compared to the narrow band images. 

The EW maps constructed with linear interpolations have a problem of mixing the thermal and non-thermal continuums, but the emission lines are only related to the thermal emission. Therefore, it is impossible to link the EW calculated this way to any physical parameters such as the abundances of different elements. Furthermore, the emission lines are sometimes not clearly isolated from the continuum, or the continuum shows significant curvature in the bands of interest. In these cases, a simple linear interpolation cannot be accurate.

\subsubsection{Continuum from spatially-resolved spectroscopy}\label{PaperIsubsubsec:2DSpecInterp}

We herein introduce a new method to construct EW maps based on our spatially-resolved spectral analysis. We first calculate the flux of the continuum, i.e., both the total flux (hereafter referred as the total continuum images) and the flux of the thermal component only (hereafter referred as the thermal continuum images), at the energy ranges of each emission lines (``Line'' in Table~\ref{PaperItable:EWmaps}) using the fitted spectrum of each tessellated regions. We have reset the abundances of all the heavy elements (O, Ne, Mg, Si, S, Ar, Ca, Fe, Ni) to \emph{zero}, so that the calculated fluxes roughly represent the underlying thermal continuum. We then use these fluxes to construct narrow band images. In order to achieve a higher spatial resolution, we renormalize the continuum flux within each region with the continuum images constructed with linear interpolations in the same band. By doing this, we have assumed this linearly-interpolated continuum images can roughly describe the surface brightness distribution of the total/thermal continuum within each tessellated regions. The total/thermal continuum images constructed this way have the same apparent spatial resolution as the original narrow band images. The EW maps are then constructed by subtracting the \emph{total} continuum images from the original narrow band images, and then being divided by the \emph{thermal} continuum images.

The EW maps constructed with this spatially-resolved spectroscopy method have relatively high spatial resolution comparable to the original narrow band images, and are related to the thermal component only. Since we have reset the abundances of all the elements to zero, small residuals in spectral fitting of the line band will not affect the EW calculation significantly, as long as other parameters of the thermal component (e.g., the normalization, temperature, and ionization parameter) are well constrained. 

We caution that the presence of heavy elements will not only affect the emission lines, but also the thermal continuum. When the metal abundances are high, the thermal continuum could even be dominated by the heavy elements instead of hydrogen and helium. Therefore, the thermal-to-non-thermal continuum ratio may change significantly by setting the abundances of heavy elements to zero and the line images could be largely contaminated by the underlying thermal continuum from heavy elements. Therefore, the EW map constructed this way should contain the contributions from both emission lines and the thermal continuum produced by heavy elements, and should be adopted as an upper limit of the true EW.

\section{Results and Discussions}\label{PaperIsec:Results}

\subsection{Spectra from larger regions}\label{PaperIsubsec:ReliabilityParaMap}

\begin{figure}[!h]
\begin{center}
\epsfig{figure=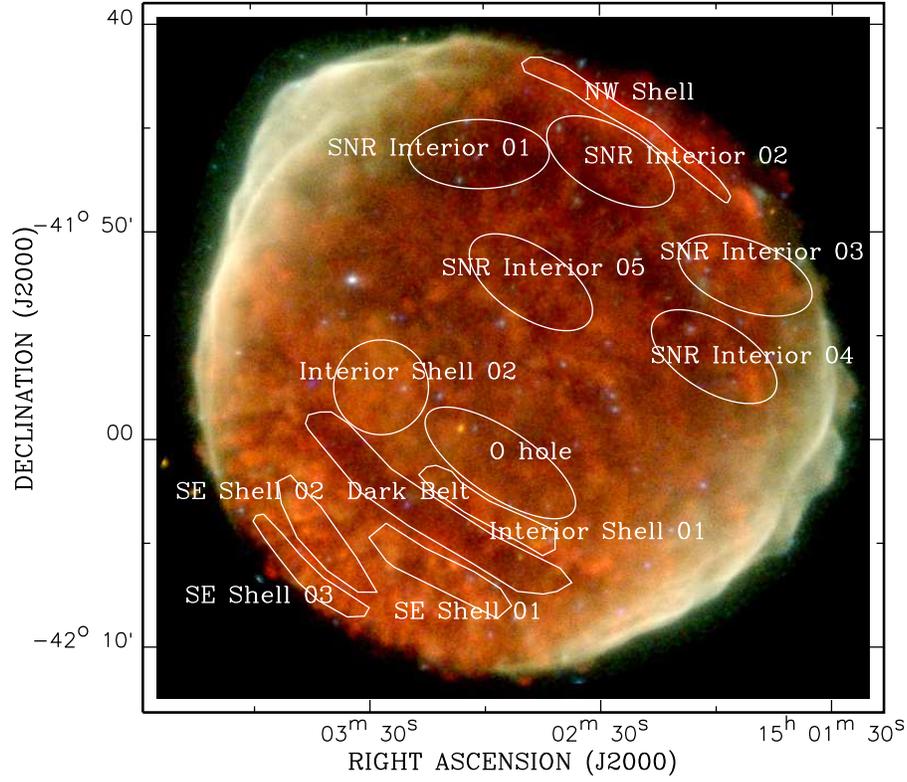,width=0.7\textwidth,angle=0, clip=}
\caption{Regions used to extract the spectra presented in Fig.~\ref{PaperIfig:specindividual} and summarized in Table~\ref{PaperItable:ParaIndividualSpec}, with region names denoted beside. The background tricolor images are the same as those presented in Fig.~\ref{PaperIfig:tricolorregions}.}\label{PaperIfig:regionsindividual}
\end{center}
\end{figure}

\begin{figure}[!h]
\begin{center}
\epsfig{figure=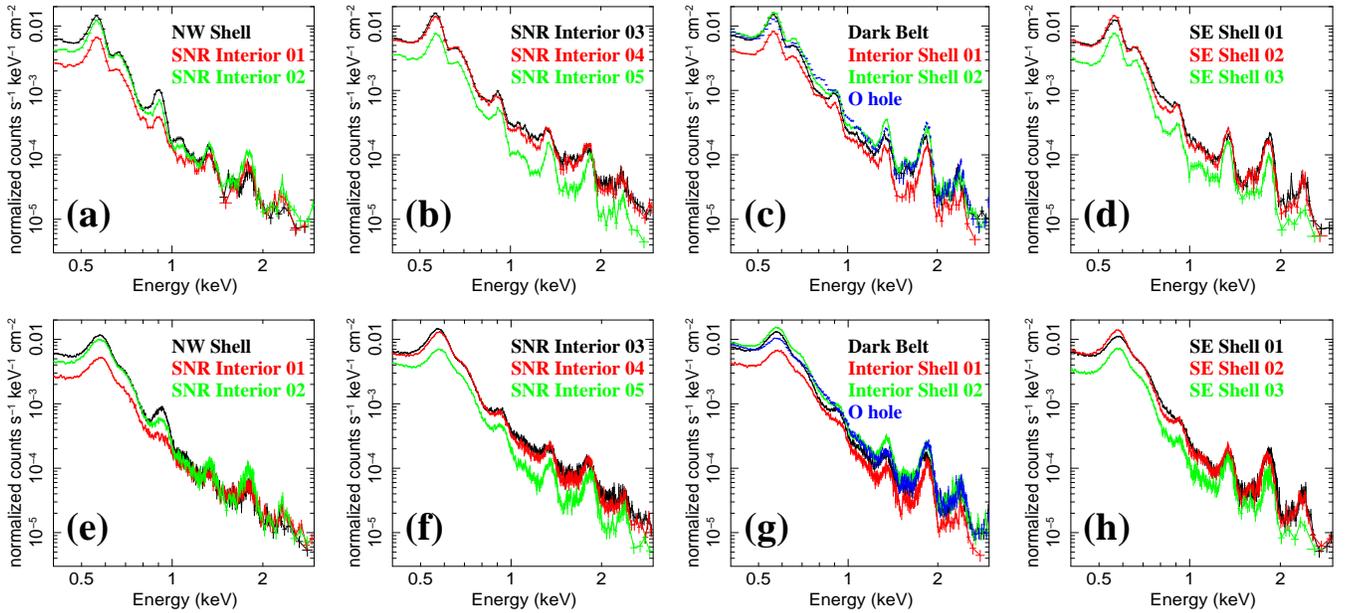,width=1.0\textwidth,angle=0, clip=}
\caption{Top row (panels a-d): the MOS-1 spectra of the 12 regions (Fig.~\ref{PaperIfig:regionsindividual}) with the region names denoted on the top right corner of each panel. We present the spectra in four different panels to clarify. Bottom row (panels e-h): similar as the top row, but for the PN spectra. MOS-2 spectra are similar as MOS-1, so are not presented here. For the convenience of comparison between the MOS and PN spectra, each spectra are divided by their effective area (in unit of $\rm counts~s^{-1}~keV^{-1}~cm^{-2}$ instead of $\rm counts~s^{-1}~keV^{-1}$). The x and y axes of each panel are in the same range.}\label{PaperIfig:specindividual}
\end{center}
\end{figure}

In order to check the reliability of parameter maps presented in the following sections, we extract spectra from larger regions (than the tessellated meshes) enclosing some interesting features. These regions are denoted in Fig.~\ref{PaperIfig:regionsindividual} and the spectra extracted from them are presented in Fig.~\ref{PaperIfig:specindividual}. We calculate the average values of some parameters of the thermal component based on the parameter images as will be presented in \S\ref{PaperIsubsec:Results2DSpec}, in order to roughly characterize the average properties of each regions. Furthermore, we also fit the spectra extracted from these larger regions using the same model as adopted for the much smaller tessellated meshes (\S\ref{PaperIsubsubsec:SpecModel}), in order to compare with the average parameters. We caution that there is generally a poor fitting of these high-counting-statistic spectra with this simple 1-T thermal plasma model, and the fitted parameters are only valid for a general comparison of the trend among different regions. The averaged parameters and the parameters obtained from spectral analysis are both summarized in Table~\ref{PaperItable:ParaIndividualSpec}.

As presented in Table~\ref{PaperItable:ParaIndividualSpec}, the fitted parameters are typically lower than the averaged parameters, especially for $\log n_et$, and the O, Ne, Mg, Si abundances. This is primarily caused by the extreme values of the parameters in some meshes with inadequately fitted spectra. As shown on the parameter maps (e.g., Fig.~\ref{PaperIfig:2DSpec_paraimg}), these extreme values are just in a few isolated meshes, but could significantly bias the average value of certain parameters in a large region. For example, the ``NW Shell'' and ``SNR Interior 01, 02'' regions have several white pixels on the $n_et$ map (extremely large $n_et$ values of $\gtrsim10^{10}\rm~cm^{-3}s$; Fig.~\ref{PaperIfig:2DSpec_paraimg}b), so their averaged $\log n_et$ show the largest differences from the fitted values. In conclusion, except for some extreme values (or ``bad pixels''), the parameter maps constructed with our spatially resolved spectroscopy method are generally consistent with the spectral analysis results of larger regions.

We use three different ways to describe the spatial distribution of emission lines from different ions: (1) the abundance map constructed with spatially resolved spectroscopy (hereafter 2-D Spec method); (2) the EW map constructed with either linear interpolation of the continuum (hereafter linear EW method) or (3) the continuum measured from the thermal component of the spatially resolved spectral modeling (hereafter 2-D Spec EW method). In principle, the 2-D Spec method is the most reliable in physics but has the lowest spatial resolution. In contrast, the linear EW method has the highest spatial resolution but cannot be directly linked to physical parameters because of the contribution from the non-thermal component and the curvature of the soft X-ray continuum (\S\ref{PaperIsubsubsec:LinearInterp}). The 2-D Spec EW method represents a compromise of these two methods, with the continuum determined from the 2-D Spec method, while the line images constructed in a similar way as for the linear EW method (\S\ref{PaperIsubsubsec:2DSpecInterp}). As the O abundance may be affected by \ion{O}{7}, \ion{O}{8}, and \ion{O}{7}~K$\delta-\zeta$ lines, we present in Table~\ref{PaperItable:ParaIndividualSpec} the EWs (from both linear EW and 2-D~Spec EW methods) of these three lines together with the O abundance. The relative strength of these lines also indicates the thermal and ionization states of the plasma.

\begin{deluxetable}{llccccccccccc}
\centering
\scriptsize 
  \tabletypesize{\scriptsize}
  \tablecaption{Average value of parameters for individual regions}
  \tablehead{
  \colhead{Region} & \colhead{Method} & \colhead{$kT$} & \colhead{$n_e$} & \colhead{$\log n_et$} & \colhead{\ion{O}{7}~$\rm K\delta-\zeta$} & \colhead{\ion{O}{7}} & \colhead{\ion{O}{8}} & \colhead{O} & \colhead{Ne} & \colhead{Mg} & \colhead{Si} & \colhead{Fe} \\
    & & keV & $\rm cm^{-3}$ & $\log(\rm cm^{-3}s)$ & EW & EW & EW & solar & solar & solar & solar & solar 
 }
\startdata
NW Shell           & Average      	& 2.25 & 0.52 & 9.56 & 0.18, 23.3 & 0.52, 12.0 & 0.33, 51.9 & 1.13 & 0.85 & 2.20 & 11.06 & 0.23 \\
		        & Fit		& 1.58 & 0.39 & 9.32 & -                & -                & -               & 0.92 & 0.64 & 1.05 & 4.84 & 0.05 \\
\hline
SNR Interior 01 & Average      	& 2.10 & 0.26 & 9.50 & 0.15, 33.1 & 0.57, 15.3 & 0.33, 61.9 & 1.77 & 0.49 & 3.81 & 32.38 & 0.18 \\
		        & Fit		& 2.22 & 0.24 & 9.37 & -                & -                & -               & 0.90 & 0.27 & 1.14 & 7.85 & 0.39 \\
\hline
SNR Interior 02 & Average      	& 1.63 & 0.33 & 9.51 & 0.21, 38.3 & 0.64, 17.8 & 0.41, 74.9 & 2.07 & 0.98 & 5.11 & 30.71 & 0.52 \\
		        & Fit		& 1.56 & 0.31 & 9.35 & -                & -                & -               & 1.05 & 0.44 & 1.69 & 12.36 & 0.83 \\
\hline
SNR Interior 03 & Average      	& 2.47 & 0.34 & 9.40 & 0.17, 23.8 & 0.60, 13.2 & 0.36, 51.5 & 1.44 & 0.50 & 3.42 & 6.95 & 0.25 \\
		        & Fit		& 4.59 & 0.28 & 9.35 & -                & -                & -               & 1.23 & 0.32 & 1.46 & 5.31 & 0.05 \\
\hline
SNR Interior 04 & Average      	& 2.87 & 0.37 & 9.48 & 0.19, 27.0 & 0.60, 10.6 & 0.44, 52.3 & 1.26 & 0.36 & 1.97 & 9.83 & 0.40 \\
		        & Fit		& 2.30 & 0.35 & 9.40 & -                & -                & -               & 1.01 & 0.28 & 1.53 & 7.36 & 0.36 \\
\hline
SNR Interior 05 & Average      	& 1.30 & 0.41 & 9.65 & 0.21, 34.8 & 0.43, 10.8 & 0.37, 59.3 & 1.53 & 0.62 & 3.16 & 20.23 & 0.18 \\
		        & Fit		& 1.12 & 0.36 & 9.60 & -                & -                & -               & 0.80 & 0.35 & 1.26 & 9.31 & 0.28 \\
\hline
Dark Belt           & Average      	& 2.06 & 0.43 & 9.49 & 0.17, 25.1 & 0.46, 8.99 & 0.33, 44.7 & 1.06 & 0.39 & 1.10 & 15.33 & 0.73 \\
		        & Fit		& 2.54 & 0.32 & 9.41 & -                & -                & -               & 0.87 & 0.32 & 0.94 & 9.30 & 0.68 \\
\hline
Interior Shell 01 & Average      	& 2.41 & 0.54 & 9.58 & 0.22, 35.9 & 0.40, 8.43 & 0.37, 50.5 & 1.06 & 0.42 & 0.86 & 9.61 & 0.99 \\
		        & Fit		& 2.20 & 0.43 & 9.53 & -                & -                & -               & 0.97 & 0.38 & 0.82 & 8.86 & 0.96 \\
\hline
Interior Shell 02 & Average      	& 2.36 & 0.53 & 9.59 & 0.20, 31.1 & 0.37, 8.85 & 0.37, 53.0 & 1.23 & 0.35 & 1.62 & 13.41 & 0.34 \\
		        & Fit		& 1.61 & 0.51 & 9.54 & -                & -                & -               & 0.77 & 0.26 & 1.01 & 8.67 & 0.44 \\
\hline
O hole               & Average      	& 1.70 & 0.51 & 9.61 & 0.16, 31.1 & 0.28, 5.74 & 0.27, 38.7 & 0.81 & 0.30 & 0.84 & 13.93 & 1.04 \\
		        & Fit		& 1.65 & 0.55 & 9.61 & -                & -                & -               & 0.43 & 0.21 & 0.47 & 5.48 & 0.63 \\
\hline
SE Shell 01       & Average      	& 1.89 & 0.53 & 9.53 & 0.17, 28.3 & 0.40, 8.38 & 0.37, 47.5 & 1.10 & 0.27 & 1.44 & 16.56 & 0.69 \\
		        & Fit		& 2.32 & 0.42 & 9.45 & -                & -                & -               & 0.99 & 0.24 & 1.32 & 12.90 & 0.76 \\
\hline
SE Shell 02       & Average      	& 1.39 & 0.46 & 9.52 & 0.16, 27.5 & 0.51, 10.0 & 0.39, 53.2 & 1.33 & 0.34 & 2.65 & 22.06 & 0.28 \\
		        & Fit		& 1.07 & 0.41 & 9.58 & -                & -                & -               & 1.23 & 0.31 & 1.82 & 17.61 & 0.18 \\
\hline
SE Shell 03       & Average      	& 1.65 & 0.32 & 9.54 & 0.22, 27.4 & 0.64, 9.63 & 0.60, 55.0 & 1.45 & 0.28 & 3.21 & 15.85 & 0.10 \\
		        & Fit		& 1.66 & 0.30 & 9.48 & -                & -                & -               & 1.15 & 0.22 & 2.24 & 11.52 & 0.05 
\enddata
\tablecomments{\scriptsize Average parameters of large regions enclosing some interesting features as denoted in Fig.~\ref{PaperIfig:regionsindividual}. For each region, the average parameters are calculated in two ways: a direct average based on the parameter images (``Average'') and the parameters obtained by fitting the MOS-1+MOS-2+PN spectra extracted from each region (e.g., Fig.~\ref{PaperIfig:specindividual}) using the model described in \S\ref{PaperIsubsubsec:SpecModel} (``Fit''). For the former method, $kT$, $\log n_et$, and $n_e$ are calculated from Fig.~\ref{PaperIfig:2DSpec_paraimg}a, b, and d. \ion{O}{7}, \ion{O}{8}, and \ion{O}{7}~$\rm K\delta-\zeta$ EWs are calculated from the linear EW maps presented in Fig.~\ref{PaperIfig:Oabundance}a-c (former numbers) and the 2-D~Spec EW maps presented in Fig.~\ref{PaperIfig:Oabundance}d-f (latter numbers). O, Ne, Mg, Si, and Fe abundances are calculated from the abundance maps in Fig.~\ref{PaperIfig:Oabundance}g, Fig.~\ref{PaperIfig:Neabundance}c, Fig.~\ref{PaperIfig:Mgabundance}c, Fig.~\ref{PaperIfig:Siabundance}c, and Fig.~\ref{PaperIfig:Feabundance}b.}\label{PaperItable:ParaIndividualSpec}
\end{deluxetable}

\subsection{Parameter maps from spatially-resolved spectroscopy}\label{PaperIsubsec:Results2DSpec}

\begin{figure}[!h]
\begin{center}
\epsfig{figure=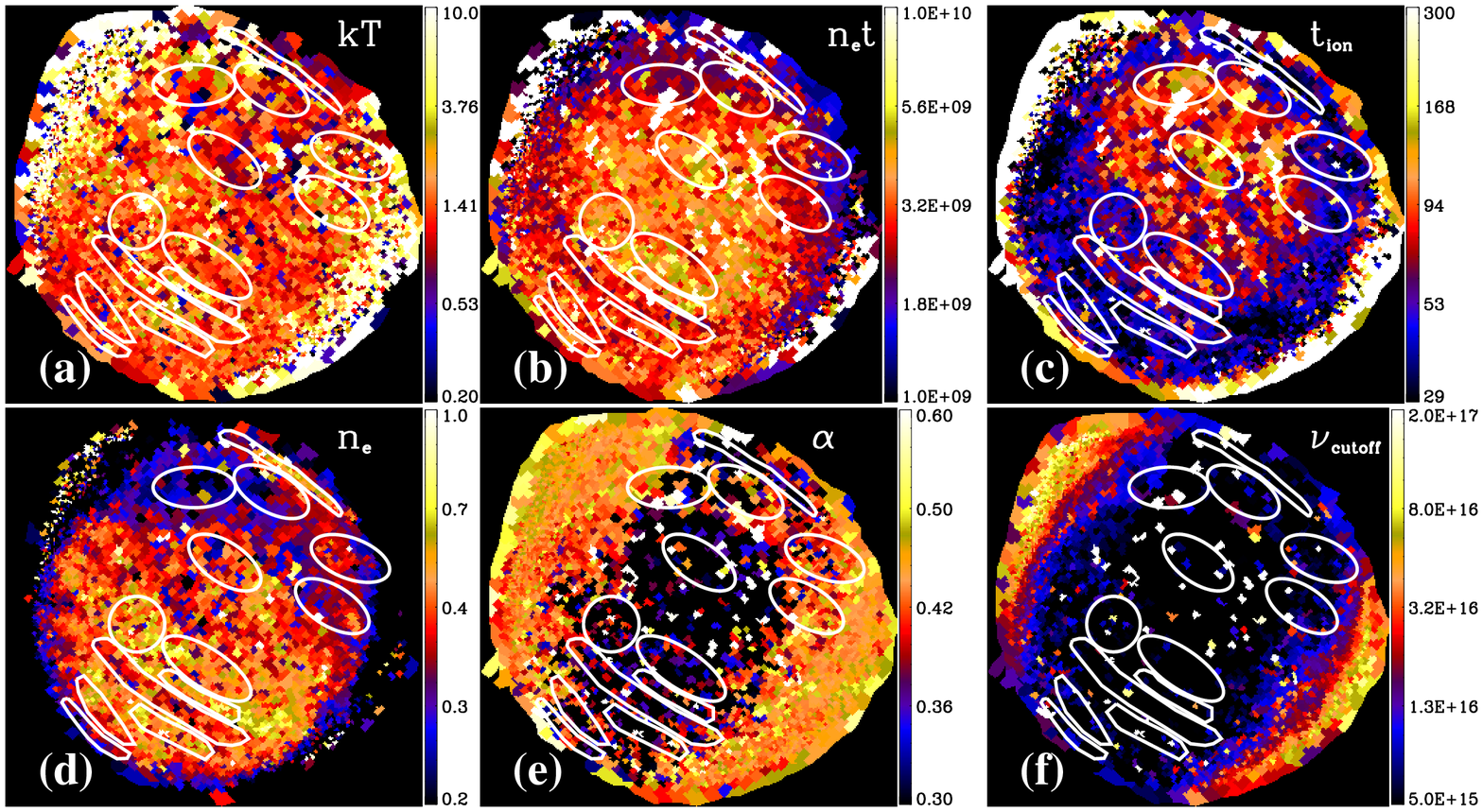,width=1.0\textwidth,angle=0, clip=}
\caption{Parameter images from spectral analysis of individual tessellated regions as shown in Fig.~\ref{PaperIfig:2DSpec_meshes}, with color bar on right and parameter name on the top right corner of each panel. (a) Temperature of the VNEI component in unit of keV ($kT$); (b) Ionization parameter of the VNEI component in unit of $\rm cm^{-3}~s$ ($n_et$); (c) Ionization time in unit of year ($t_{ion}$); (d) Electron number density in unit of $\rm cm^{-3}$ ($n_e$); (e) Radio spectral index of the SRCUT component ($\alpha$); (f) Cutoff frequency of the SRCUT component ($\nu_{cutoff}$). White regions overlaid on each panel are denoted in Fig.~\ref{PaperIfig:regionsindividual} and described in \S\ref{PaperIsubsec:ReliabilityParaMap}.}\label{PaperIfig:2DSpec_paraimg}
\end{center}
\end{figure}

We present the parameter maps constructed with the spatially resolved spectroscopy method in Fig.~\ref{PaperIfig:2DSpec_paraimg}. In the following sections, we describe interesting structures on each map and discuss their physical implications. 

\subsubsection{Thermal and ionization states}\label{PaperIsubsubsec:2DSpec_kT}

The thermal and ionization states (characterized by $kT$ and $n_et$) trace the shock heating and ionization history of the SNR. In young SNRs, the forward and reverse shocked gas is often far from temperature and ionization equilibrium (e.g., \citealt{Slane14}). X-ray measurements in the literatures reveal a large diverse of $kT$ and $n_et$ in SN1006 (with either 1-T model or more complicated models), with $kT$ ranging from $\lesssim0.5\rm~keV$ to $\gtrsim 4\rm~keV$, and $n_et$ typically in the range of a~few$\times10^9\rm~cm^{-3}s$ (e.g., \citealt{Koyama95,Vink00,Dyer01,Allen01,Vink03,Long03,Acero07,Yamaguchi08,Miceli12,Uchida13}). Although the presence of non-thermal emission in the NE and SW synchrotron rims may lead to large systematic uncertainties in the derived $kT$ and $n_et$, which may partly account for the spread in the reported $kT$ and $n_et$ measurements, for the interior of the SNR the range of measured $kT$ and $n_et$ likely reflect intrinsic temperature and ionization state variations.

In the present paper, we adopt a simple 1-T model to characterize the thermal plasma emission (\S\ref{PaperIsubsubsec:SpecModel}). $kT$ and $n_et$ are thus highly affected by the most prominent O lines, as well as the Ne, Mg, and Fe~L lines, but not significantly affected by the poorly resolved Si and S lines. We present the $kT$ and $n_et$ maps in Fig.~\ref{PaperIfig:2DSpec_paraimg}a,b, and caution that $kT$ and $n_et$ in the NE and SW non-thermal limbs are not well constrained and could be significantly overestimated in some regions. 

In the regions dominated by thermal emission, the thermal and ionization states of the plasma show clear spatial variations. $kT$ typically varies in the range of $\sim1-4\rm~keV$, while $n_et$ typically varies in the range of $10^{9-10}\rm~cm^{-3}s$. These ranges are generally consistent with previous measurements. The outer shells of the SNR (the NW and SE shells) appear to have low $kT$ and $n_et$, indicating that the gas there are newly shocked and are far from thermal and ionization equilibrium. $kT$ of the SNR interior is relatively smoothly distributed, with most of the regions having $kT\sim1.3\rm~keV$. Some higher temperature structures are superimposed, such as ``SNR Interior 03 and 04'', ``Interior Shell 01 and 02'', etc. These features are typically locally high-density structures. In contrast, the $n_et$ map shows clear gradient in spatial distribution, with the highest $n_et$ appears close to the center of the SNR (``O hole''), and declines outward. The ``Dark Belt'' has clearly lower $n_et$ than the surrounding regions.

\begin{figure}[!h]
\begin{center}
\epsfig{figure=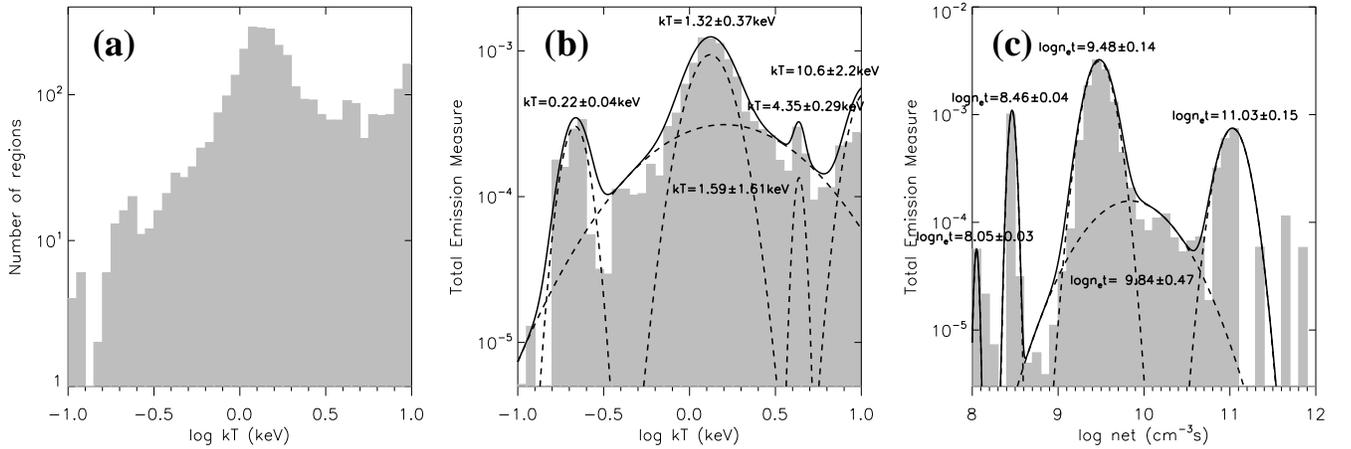,width=1.0\textwidth,angle=0, clip=}
\caption{Probability distribution functions (PDFs) of the temperature ($kT$) and ionization parameter ($n_et$) of the VNEI component of all the 3596 tessellated regions. Panel~(a) is the PDF of $kT$ with the y-axis denoting the number of regions, while panels (b) and (c) are the emission-measure-weighted PDFs with the y-axis denoting the total emission measure. The solid curve in panels~(b) and (c) are the best-fit multi-Gaussian model, with the dashed curves mark the model components. The central positions of each Gaussian component are also denoted. The errors are at 1-$\sigma$ level, obtained from the width of the Gaussian component.}\label{PaperIfig:PDF}
\end{center}
\end{figure}

We characterize the distribution of the thermal and ionization states with the original and emission-measure-weighted (EM-weighted) probability distribution functions (PDFs) (Fig.~\ref{PaperIfig:PDF}). The EM-weighted PDFs of both $kT$ and $n_et$ can be characterized with a 5-Gaussian model. Interestingly, the three peaks at 0.22~keV, 1.32~keV (or the broad Gaussian centered at 1.59~keV), and 4.35~keV in the $kT$ PDF are roughly consistent with the three components from the spectral analysis of \citet{Uchida13}, where they attribute the lowest temperature component to shocked ISM, while the other two components to shocked ejecta. Nevertheless, as shown in Fig.~\ref{PaperIfig:PDF}b,c, three of the five Gaussians are relatively narrow and have low EM (the $kT=0.22$, 4.35, and 10.6~keV components); they are either in the non-thermal emission dominated regions so have poorly constrained $kT$ and $n_et$ or are in some small isolated regions in the SNR interior. We therefore only use the two most prominent Gaussians to characterize the average thermal and ionization states of SN1006. We assume the $kT=1.32\pm0.37\rm~keV$ component and the $\log n_et/({\rm cm^{-3}s})=9.48\pm0.14$ component are from the same regions, while the $kT=1.59\pm1.61\rm~keV$ component and the $\log n_et/({\rm cm^{-3}s})=9.84\pm0.47$ component are from the other regions.

\begin{figure}[!h]
\begin{center}
\epsfig{figure=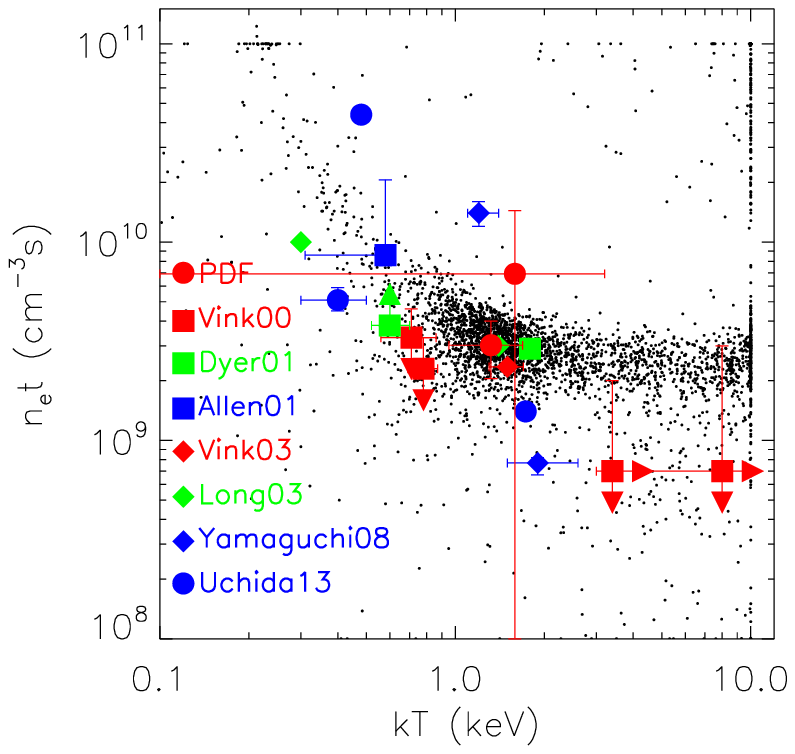,width=0.67\textwidth,angle=0, clip=}
\caption{$kT$ v.s. $n_et$ for all the 3596 tesselllated regions (black dots). The red filled circles (``PDF'') are the two most prominent Gaussian components in Fig.~\ref{PaperIfig:PDF}b,c: the strongest component with $kT=1.32\pm0.37\rm~keV$ and $\log n_et/({\rm cm^{-3}s})=9.48\pm0.14$ and the broadest component with $kT=1.59\pm1.61\rm~keV$ and $\log n_et/({\rm cm^{-3}s})=9.84\pm0.47$. Other colored symbols are archival measurements of SN1006 with the references denoted in the lower left corner.\\
\smallskip\\
References: Vink00: \citet{Vink00}; Dyer01:\citet{Dyer01}; Allen01: \citet{Allen01}; Vink03: \citet{Vink03}; Long03: \citet{Long03}; Yamaguchi08: \citet{Yamaguchi08}; Uchida13: \citet{Uchida13}.}\label{PaperIfig:kTnet}
\end{center}
\end{figure}

We further compare our measured $kT$ and $n_et$ of all the tessellated regions to those obtained from the archival measurements of SN1006 (Fig.~\ref{PaperIfig:kTnet}). We also plot the two most prominent components in the PDFs of $kT$ and $n_et$ to represent the average thermal and ionization states of SN1006. The archival measurements are obtained with different data, from different regions, and/or with different models, so span a large range of $kT$ and $n_et$. Nevertheless, most of these measurements are roughly consistent with our measurements. Therefore, the large diversity in the measured $kT$ and $n_et$ from previous works may simply because they are obtained in different regions consisting of plasma at various thermal and ionization states.

Combining the $n_e$ and $n_et$ maps, we construct the ionization age ($t_{ion}$) map with truly time dimension (in unit of year; Fig.~\ref{PaperIfig:2DSpec_paraimg}c). Except for the bright rim surrounding the SNR, which is artificial due to the low flux density of the surrounding regions, the whole SNR shell appears to have a low and smooth ionization age of $t_{ion}\lesssim150\rm~year$. In contrast, all the regions in the SNR interior have $t_{ion}>150\rm~year$, with the highest $t_{ion}\sim500\rm~year$, consistent with the age of SN1006 of $\sim10^3\rm~year$. The $t_{ion}$ map suggests that the plasma in the SNR shell is probably newly shocked, while the SNR interior was shocked several hundred years before after the SN explosion and is on the approach of ionization equilibrium. 

\subsubsection{Electron density}\label{PaperIsubsubsec:2DSpec_ne}

Many previous studies indicate that SN1006 locates in a low-density environment. The measured density of the surrounding ISM is typically in the range of $0.04\lesssim n_0 \lesssim0.4\rm~cm^{-3}$ and the NW part is evidenced to have the highest density. Different estimates of the ISM density are based on \ion{H}{1} \citep{Dubner02}, IR \citep{Winkler13}, Balmer lines \citep{Raymond07,Nikolic13}, and UV \citep{Laming96,Korreck04} observations, as well as the thermal component of the X-ray emission \citep{Winkler97,Long03,Acero07,Miceli12}. This low environmental density also explains SN1006's faintness in TeV $\gamma$-ray emission, which could partly come from the proton-proton interaction with $\pi^0$ production and subsequent decay \citep{Aharonian05,Acero10,Acero15}.

For the first time, we obtain the map of electron number density of the thermal plasma ($n_e$) over the entire SNR. As shown in Fig.~\ref{PaperIfig:2DSpec_paraimg}d, the overall $n_e$ distribution resembles the total surface brightness distribution, except for the two non-thermal limbs. The $n_e$ image also resembles the ``thermal'' image presented in \citet{Miceli09}, consistent with its origin from the thermal emission. The ``NW shell'' has significantly higher $n_e$ than the ``SNR Interior''. Assuming no CR acceleration in this part so a compression ratio of 4, the ambient ISM density surrounding the ``NW shell'' should be $n_0\sim0.15\rm~cm^{-3}$, consistent with previous multi-wavelength estimates. There is no sharp increase of $n_e$ in the SE part. Instead, the ``SE Shell'' appears much thicker than the ``NW Shell'', and the SE half of the SNR is separated by the low density ``Dark Belt'' into two shells (the ``SE Shell'' and the ``Interior Shell''), both with comparable $n_e$ as the ``NW Shell''. The ``Dark Belt'' has low $n_e$ and $n_et$ compared to the surrounding regions (``Interior Shell'', ``SE Shell 01''), but has $t_{ion}$ comparable to or even higher than these regions. The low ionization states of the ions in the ``Dark Belt'' are further indicated by its relatively low \ion{O}{8}/\ion{O}{7} ratio (Fig.~\ref{PaperIfig:specindividual}c; Table~\ref{PaperItable:ParaIndividualSpec}) and low centroid energy of the \ion{Si}{13} lines (Fig.~\ref{PaperIfig:specindividual}c,g). The outermost part of the ``SE Shell'' (e.g., ``SE Shell 03'') decreases in $n_e$. The northern part of the ``SNR Interior'' has a clearly lower density of $n_e\lesssim0.3\rm~cm^{-3}$, but ``SNR Interior 03 and 04'' may form a shell-like structure behind the SW non-thermal rim, apparently extending the ``NW Shell''. In higher resolution intensity images (e.g., Fig.~\ref{PaperIfig:regionsindividual}), many parts of the SNR interior, such as the ``SE Shell'' and the ``Interior Shell'', are resolved into clumpy structures, in contrast to the significantly filamentary structure in the ``NW Shell''.

Based on the $n_e$ map, we could further estimate the total mass of the plasma contained in the 3596 tessellated regions, again assuming a solar abundance. The estimated mass of the shocked X-ray emitting plasma is $\sim14.5f^{\frac{1}{2}}\rm~M_\odot$, where $f$ is the volume filling factor. As discussed in \S\ref{PaperIsubsubsec:DerivedPara}, the swept up ambient ISM mass is $\sim5\rm~M_\odot$. Adding the mass of the shocked ejecta, which is quite uncertain but contribute only a small fraction to the mass budget, we could roughly constrain the volume filling factor to be $f\sim0.1$ under the adopted geometric model (\S\ref{PaperIsubsubsec:DerivedPara}).

\subsubsection{Non-thermal component}\label{PaperIsubsubsec:2DSpec_non-thermalcomponent}

The shape of the non-thermal X-ray spectra is an indicator of many parameters of particle acceleration at the SNR shocks (e.g., \citealt{Allen01,Allen08,Rothenflug04,Miceli09,Miceli13}). We will further discuss the modeling of the non-thermal spectrum of SN1006 in companion papers, including our newly approved \emph{NuSTAR} Cycle~1 observations of the non-thermal rims of SN1006 (400~ks; PI: Li Jiang-Tao). We herein present the preliminary results on the non-thermal X-ray emission from the simplest spectral decomposition with the ``VNEI+SRCUT'' model. As described in \S\ref{PaperIsubsubsec:SpecModel}, we fix the normalization of the SRCUT component using the 1.4~GHz flux-accurate image of SN1006 from \citet{Dyer09}. Therefore, the only two free parameters of the SRCUT component are the radio-to-X-ray photon index $\alpha$ and the cutoff frequency $\nu_{cutoff}$, which are presented in Fig.~\ref{PaperIfig:2DSpec_paraimg}e,f.

$\alpha$ of the regions dominated by non-thermal emission is quite smooth, roughly in the range of 0.4-0.5. This is consistent with the adoption of a constant $\alpha\sim0.45-0.6$ in some previous works (e.g., \citealt{Dyer01,Acero07,Katsuda10}). However, it is also possible that there is a more complicated shape of the non-thermal X-ray emission and $\alpha$ is varying not only at different locations, but also at different energies (e.g., \citealt{Allen08,Miceli13}). The current energy coverage (with \emph{Chandra} or \emph{XMM-Newton}) is too narrow in hard X-ray and we expect a better modeling of the electron synchrotron emission with our new \emph{NuSTAR} data.

The cutoff frequency $\nu_{cutoff}$ shows significant spatial variations within the non-thermal emission dominated regions, as previously revealed by \citet{Rothenflug04,Miceli09}. The bright non-thermal filaments are resolved on the $\nu_{cutoff}$ map and have higher $\nu_{cutoff}$ of $>5\times10^{16}\rm~Hz$, while the inter-filament regions typically have $\nu_{cutoff}\sim(2-5)\times10^{16}\rm~Hz$. $\nu_{cutoff}$ continuously falls toward the SNR interior, and is $<5\times10^{15}\rm~Hz$ in the regions dominated by thermal emission.

\subsection{Spatial distribution of metals}\label{PaperIsubsec:Abundance}

Although the presence of emission lines from metal ions in SN1006 has been suggested in some early X-ray observations \citep{Becker80,Leahy91}, these lines are not reliably detected until the publication of \emph{ASCA} data \citep{Koyama95}. Later on, with \emph{BeppoSAX}, \emph{Chandra}, \emph{XMM-Newton}, and \emph{Suzaku} observations, the emission lines are better resolved and the spatial distributions of metal ions have been extensively studied, either with spectral analysis of different regions or the narrow band mapping of different emission lines (e.g., \citealt{Vink00,Vink03,Long03,Acero07,Yamaguchi08,Uchida13}). We herein study the spatial distribution of different elements in SN1006, using both the abundance map and the EW maps constructed with the linear EW or 2-D Spec EW method.

\subsubsection{O}\label{PaperIsubsubsec:metal_Oline}

\begin{figure}[!h]
\begin{center}
\epsfig{figure=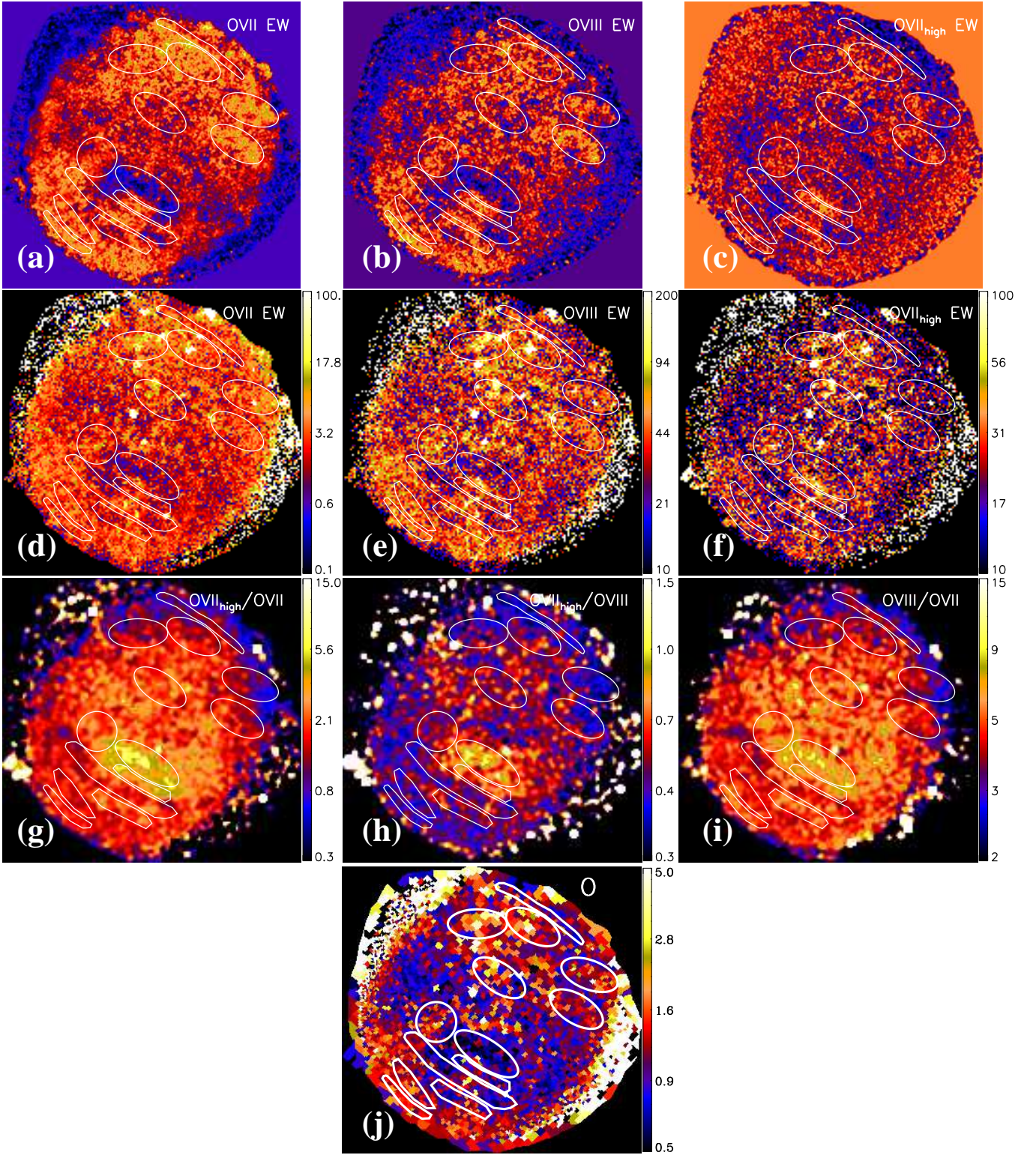,width=1.0\textwidth,angle=0, clip=}
\caption{(a-c) \ion{O}{7}, \ion{O}{8}, and \ion{O}{7}~$\rm K\delta-\zeta$ EW maps constructed with the linear EW method. Background pixels are set to 0, so pixels with color bluer than the background color have negative values. (d-f) Similar as panels (a-c), but constructed with the 2-D Spec EW method. (g-i) The EW ratio of \ion{O}{7}~$\rm K\delta-\zeta$/\ion{O}{7}, \ion{O}{7}~$\rm K\delta-\zeta$/\ion{O}{8}, and \ion{O}{8}/\ion{O}{7} constructed with the 2-D Spec EW method. (j) Oxygen abundance map obtained by fitting the spectra in each tessellated regions, with the color bar in unit of solar abundance. White regions overlaid on each panel are denoted in Fig.~\ref{PaperIfig:regionsindividual} and described in \S\ref{PaperIsubsec:ReliabilityParaMap}.}\label{PaperIfig:Oabundance}
\end{center}
\end{figure}

There are usually two prominent oxygen line bumps presented in the soft X-ray spectrum of SN1006: the \ion{O}{7} K-shell transitions at $\sim0.56-0.58\rm~keV$ and the \ion{O}{8} K$\beta$ and K$\gamma$ lines at $\sim0.65\rm~keV$, and sometimes a third line at $\sim0.7-0.75\rm~keV$ representing the \ion{O}{7} K$\delta-\zeta$ transitions (Fig.~\ref{PaperIfig:EWband}). In Fig.~\ref{PaperIfig:Oabundance}, we present the \ion{O}{7}, \ion{O}{8}, and \ion{O}{7} K$\delta-\zeta$ EW maps constructed with both the linear EW method and the 2-D Spec EW method, the 2-D Spec EW ratios of these lines, as well as the O abundance map. 

As shown in Fig.~\ref{PaperIfig:Oabundance}, the intensities of \ion{O}{7} and \ion{O}{8} gradually decrease toward the center, with the lowest intensity at the ``O Hole''. This spatial variation indicates a gradual decrease of the O abundance toward the center. The O abundance across the whole SNR is close to solar value, except for the non-thermal arcs where the properties of the thermal component are not well constrained. Therefore, expect for some fine structures, such as the ``SE Shell 02 and 03'' (also refer to Table~\ref{PaperItable:ParaIndividualSpec}), we generally do not find significant evidence of O enrichment in SN1006. 

The \ion{O}{7} and \ion{O}{8} distributions show some noticeable differences. As shown on the \ion{O}{8}/\ion{O}{7} map (Fig.~\ref{PaperIfig:Oabundance}i), in the NW edge, the \ion{O}{7} lines appear to be relatively stronger, while the \ion{O}{8} lines are stronger in the opposite side (the SE half of the SNR). The \ion{O}{8}/\ion{O}{7} map resembles the $n_et$ map (Fig.~\ref{PaperIfig:2DSpec_paraimg}b), indicating that $n_et$ is primarily determined by the most prominent O lines. As revealed by the \ion{O}{8}/\ion{O}{7} and $n_et$ maps, the plasma in the SE half of SN1006 is closer to ionization equilibrium. However, this higher $n_et$ is most likely a result of the higher density in this region, as the ionization time ($t_{ion}$) at the SE and NW edges seem to be comparable (Fig.~\ref{PaperIfig:2DSpec_paraimg}c).

The only feature with significantly subsolar O abundance is the ``O Hole'' (Table~\ref{PaperItable:ParaIndividualSpec}). It is interesting that this region, in addition to the ``Interior Shell 01'' close to it, also has significantly enhanced \ion{O}{7}~$\rm K\delta-\zeta$/\ion{O}{7} and \ion{O}{7}~$\rm K\delta-\zeta$/\ion{O}{8} ratios (Fig.~\ref{PaperIfig:Oabundance}g,h). The \ion{O}{7}~$\rm K\delta-\zeta$/\ion{O}{7} map is also very similar in morphology as the $n_et$ map, indicating that the regions with strong \ion{O}{7}~$\rm K\delta-\zeta$ emission apparently have high ionization state. However, this may instead be an artifact because it is easy to mix \ion{O}{7}~$\rm K\delta-\zeta$ and \ion{O}{8} lines. Alternatively, the  \ion{O}{7}~$\rm K\delta-\zeta$ lines may also be highly contaminated by the low ionization transitions of Fe~L lines. We will further discuss this possibility in \S\ref{PaperIsubsubsec:metal_FeK}.

\subsubsection{Ne}\label{PaperIsubsubsec:metal_Ne}

In case of no prominent iron L-shell lines, the significant emission line bump at $\sim0.9\rm~keV$ mainly consists of \ion{Ne}{9} K-shell transitions (Fig.~\ref{PaperIfig:EWband}, Table~\ref{PaperItable:EWmaps}). Some regions with strong Ne emissions also show weak \ion{Ne}{10} K-shell emission lines (at $\sim1.0\rm~keV$; e.g., the ``NW Shell''; Fig.~\ref{PaperIfig:specindividual}a), which are much less significant in the spectra of other regions. 

\begin{figure}[!h]
\begin{center}
\epsfig{figure=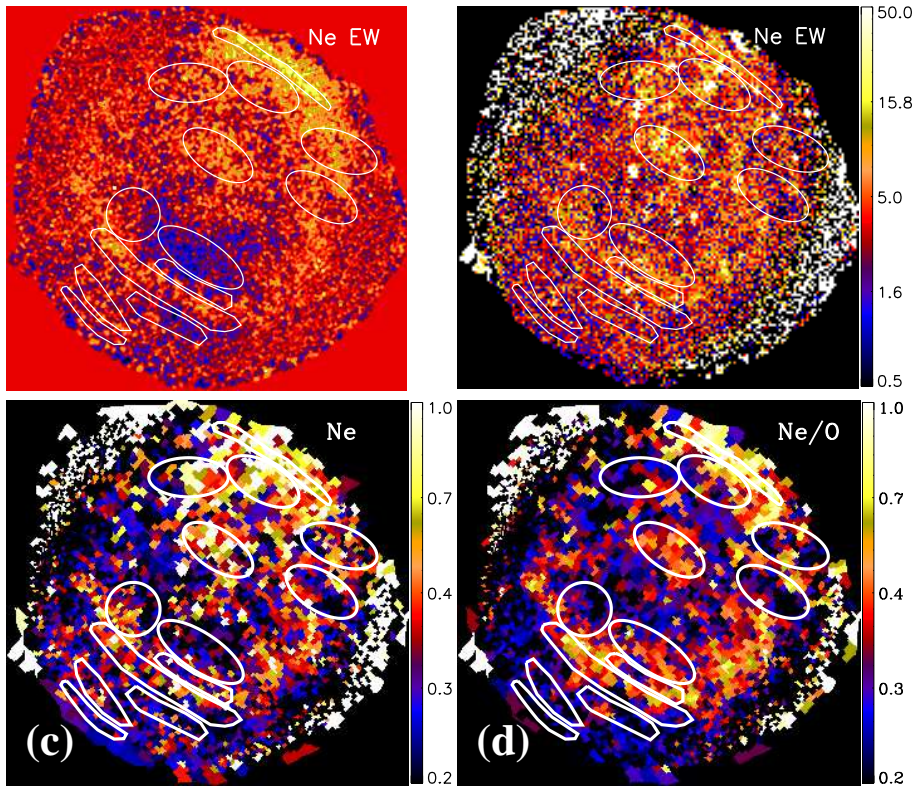,width=0.67\textwidth,angle=0, clip=}
\caption{The \ion{Ne}{9} EW maps constructed with (a) the linear EW method and (b) the 2-D Spec EW method. (c) The Ne abundance map obtained by fitting the spectra in each tessellated regions. (d) Ne/O abundance ratio map.}\label{PaperIfig:Neabundance}
\end{center}
\end{figure}

As shown in Fig.~\ref{PaperIfig:Neabundance}, the Ne abundance map (panel~c and d) shows strikingly similar features as the Ne EW maps (panels a and b). Some features with enhanced Ne emission are clearly resolved, such as the ``NW Shell'', the weak enhancement roughly at the ``Dark Belt'' (or a little behind, i.e., toward NW) and some parts of the ``SNR Interior''. In contrast, the ``SE Shell'' has extremely low Ne emission. The coincidence of these features in different panels of Fig.~\ref{PaperIfig:Neabundance} strongly suggest the enhancement at the NW shell is related to the SNR itself, instead of just an artificial background feature.

Except for the ``NW Shell'' ($\rm Ne/O\gtrsim0.7$; Table~\ref{PaperItable:ParaIndividualSpec}), the Ne/O ratio is significantly below solar across the entire SNR (Fig.~\ref{PaperIfig:Neabundance}d). The highest Ne/O ratio inside the SNR is $\lesssim0.5$ (at the ``Dark Belt'' and ``O Hole'') and the typical value of the Ne/O ratio is $\sim$0.2-0.3. The NW high-Ne filament also has high electron density, as well as strong soft X-ray and H$\alpha$ emissions, which are often signatures of the shocked ISM (e.g., \citealt{Winkler14}). Therefore, Ne in SN1006 is most likely from the ISM instead of the ejecta, consistent with the small amount of Ne yielded in Type~Ia SN \citep{Iwamoto99}. If the high-Ne regions represent shocked ISM, it is very likely that a significant fraction of the O line emissions are from the SN ejecta, although the O abundance is mostly solar across the SNR (\S\ref{PaperIsubsubsec:metal_Oline}), as the O lines are always weak in high-Ne regions.

\subsubsection{Mg}\label{PaperIsubsubsec:metal_Mg}

\begin{figure}[!h]
\begin{center}
\epsfig{figure=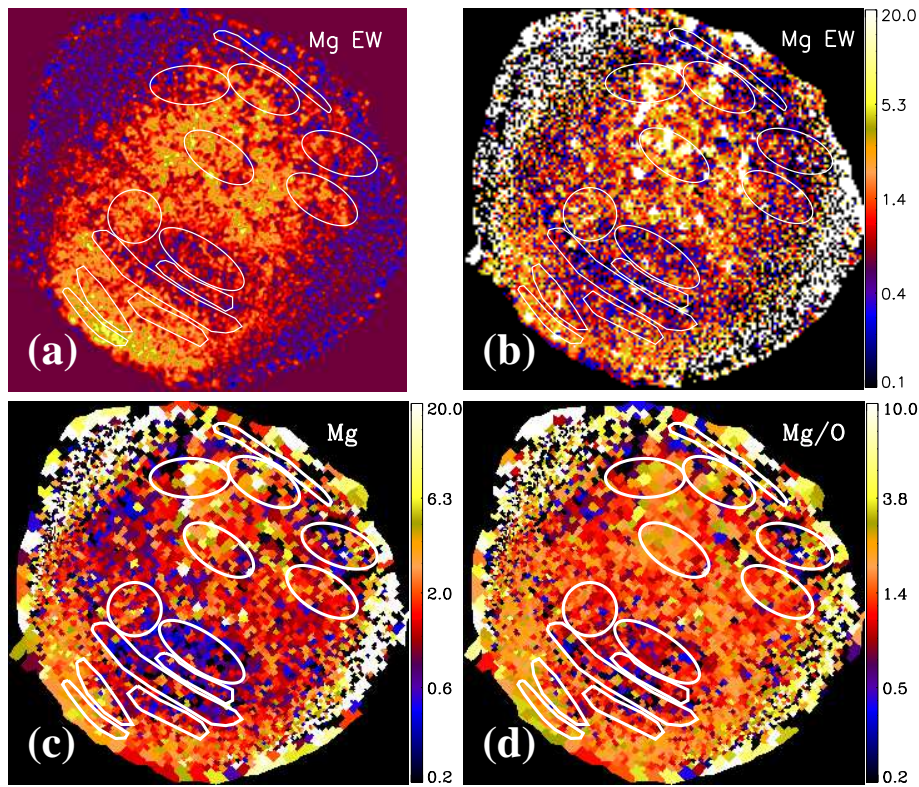,width=0.67\textwidth,angle=0, clip=}
\caption{Similar as Fig.~\ref{PaperIfig:Neabundance}, but for \ion{Mg}{11} lines and the Mg abundance.}\label{PaperIfig:Mgabundance}
\end{center}
\end{figure}

In case of no prominent iron L-shell lines, the significant line bump at $\sim1.3\rm~keV$ mainly consists of \ion{Mg}{11} K-shell transitions and possibly some weak \ion{Ne}{10} K-shell lines (Fig.~\ref{PaperIfig:EWband}, Table~\ref{PaperItable:EWmaps}). Similarly, the Mg abundance map also shows coincident features to the Mg EW maps constructed with either the linear EW or the 2-D Spec EW method (Fig.~\ref{PaperIfig:Mgabundance}), such as the low intensity at the ``O Hole'', ``Interior Shell 01'', ``Dark Belt'' and the relatively high intensity in the ``SNR Interior'' and ``SE Shell''. Type~Ia SNe are not expected to yield a lot of Mg \citep{Iwamoto99}, so Mg should be mostly ISM in origin. However, the spatial distributions of Mg and Ne are quite different. For example, compared to the Ne lines, the Mg line strength is pretty low at the ``NW Shell'' but high at the ``SE Shell'' (Fig.~\ref{PaperIfig:specindividual}; Table~\ref{PaperItable:ParaIndividualSpec}). These differences in Ne and Mg distributions indicate that either SN1006 is Mg enriched due to some unknown reasons or it is more likely that the ISM conditions (density, ionization state, etc.) are significantly variable around SN1006, which produces variable Mg/Ne emission line ratios even though their abundance ratio is not significantly variable. In most of the regions, the Mg/O ratio is $>1$ (Fig.~\ref{PaperIfig:Mgabundance}d). However, this is very likely to be an artificial effect as we have adopted a 1-T model which often fit the most prominent O lines well so may have overpredicted the abundances and EWs of emission lines at higher energy.

\subsubsection{Si and S}\label{PaperIsubsubsec:metal_SiS}

Because of the low counting statistics, Si lines are often weak in the spectra of individual tessellated regions, and S lines are even weaker. We therefore link the Si and S abundances in spectral fitting, by assuming that they are produced in related processes in stellar nuclear synthesis (e.g., oxygen burning) so are co-spatial. There are two prominent Si line bumps in the spectra of SN1006 (Fig.~\ref{PaperIfig:EWband}). The \ion{Si}{13} bump centers at $\sim1.8\rm~keV$ and mainly consists of K-shell transitions of helium-like silicon ions, although some higher level transitions and \ion{Mg}{12} K-shell lines may also contribute. There is also an emission line feature centered at $\sim2.1\rm~keV$, and is most likely blueshifted \ion{Si}{14} K$\beta$ and K$\gamma$ lines with a rest frame energy of $\sim2.0\rm~keV$. Another weak bump centered at 2.4-2.5~keV is most likely \ion{S}{15} K-shell transitions, although some weak \ion{Si}{14} or even \ion{Si}{13} K-shell transitions may contribute.

\begin{figure}[!h]
\begin{center}
\epsfig{figure=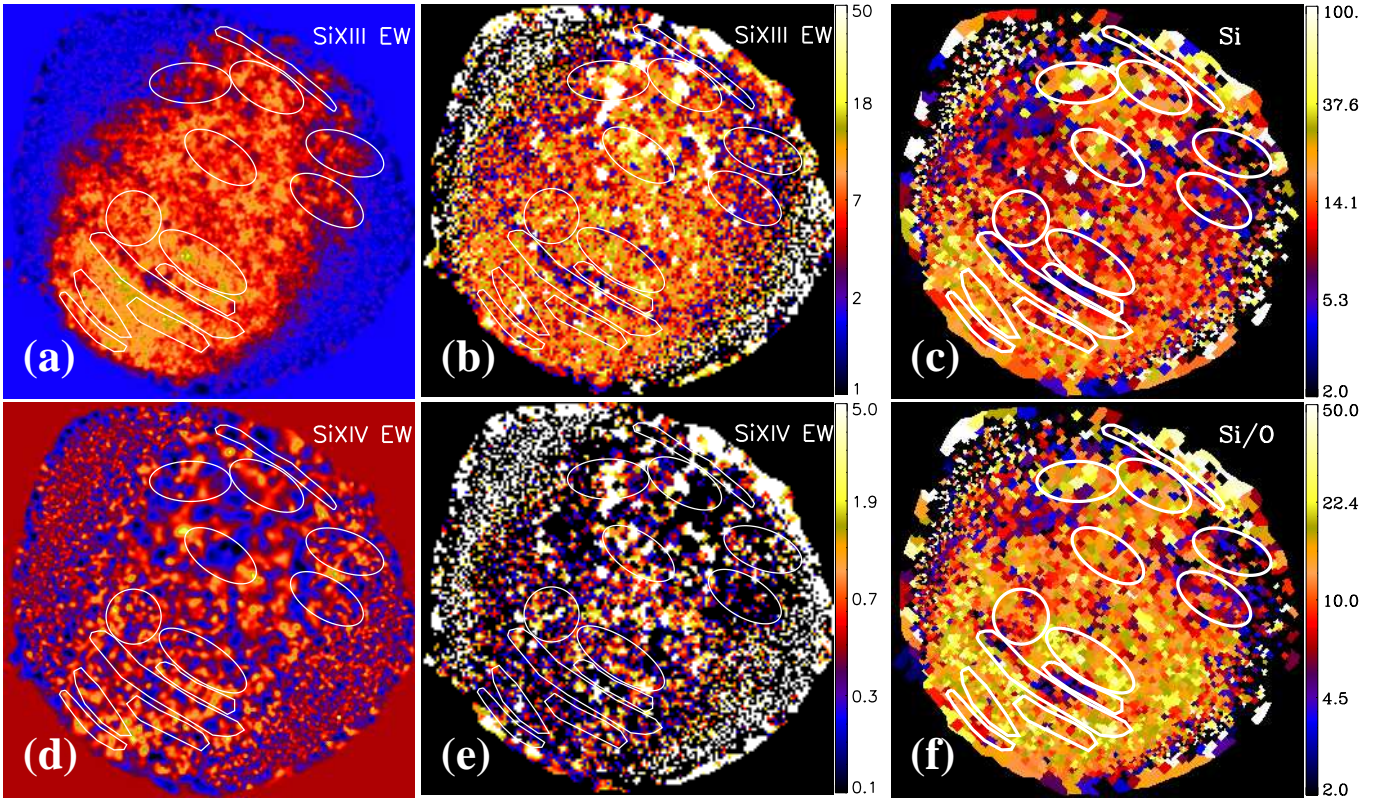,width=1.0\textwidth,angle=0, clip=}
\caption{Panels (a-c) and (f) are similar as Fig.~\ref{PaperIfig:Neabundance}, but for \ion{Si}{13} lines and the Si abundance. Panels (d, e): the \ion{Si}{14} EW maps constructed with the linear EW method and the 2-D Spec EW method.}\label{PaperIfig:Siabundance}
\end{center}
\end{figure}

The \ion{Si}{13} and \ion{Si}{14} EW maps and the Si abundance map are presented in Fig.~\ref{PaperIfig:Siabundance}. The \ion{Si}{14} lines are very weak and have no detectable high signal-to-noise ratio features on the images, although some enhancements in the SNR interior may still be possible. On the other hand, the \ion{Si}{13} lines are quite prominent and determine the measurement of Si abundance. The \ion{Si}{13} line bump is the strongest in the ``SE Shell'' and shows clearly asymmetric distribution. The outermost shells (``NW Shell'' and ``SE Shell 03''), as well as the regions slightly NW to the ``Dark Belt'', appear to be relatively weak in \ion{Si}{13} emission. The Si/O ratio is above solar (Fig.~\ref{PaperIfig:Siabundance}f). Similar as the Mg/O ratio, it may also be an artificial effect caused by the 1-T thermal plasma model.

``SNR Interior 01-05'' represent regions selected from the low surface brightness interior between the NW shell and the relatively bright SE half. The five regions have the same surface area, so their spectra shown in Fig.~\ref{PaperIfig:specindividual}a,b and e,f are also indicative of the intrinsic intensity of each region. It is clear that ``SNR Interior 01, 02, and 05'' are fainter than the other two, which apparently coincide with two shell-like features projected just behind the west non-thermal shell. ``SNR Interior 01'' has the lowest electron density, consistent with its low surface brightness. Compared to ``SNR Interior 01, 02, and 05'', the Si abundances (or the Si/O ratio) of ``SNR Interior 03 and 04'' are lower (Table~\ref{PaperItable:ParaIndividualSpec}), as also indicated by their weaker Si lines (Fig.~\ref{PaperIfig:specindividual}b,f). It is therefore likely that ``SNR Interior 03 and 04'' are largely comprised of shocked ISM if the Si lines are mostly from the SN ejecta.

\begin{figure}[!h]
\begin{center}
\epsfig{figure=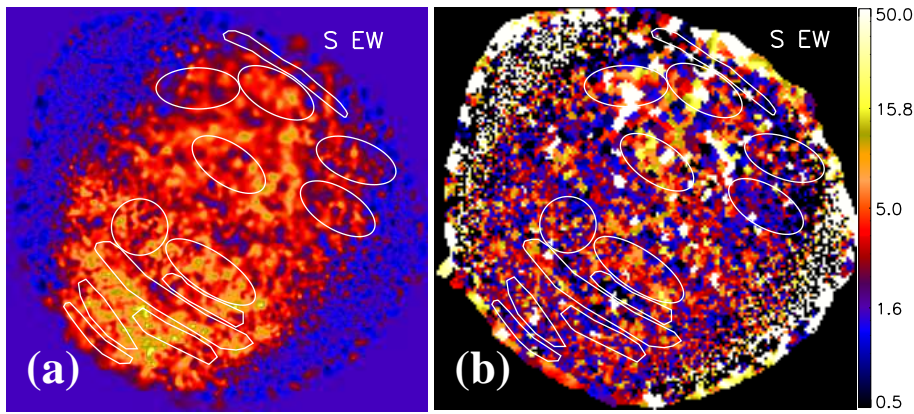,width=0.67\textwidth,angle=0, clip=}
\caption{The \ion{S}{15} EW maps constructed with (a) the linear EW method and (b) the 2-D Spec EW method.}\label{PaperIfig:Sabundance}
\end{center}
\end{figure}

In order to study the possibly different distributions of Si and S, we also present the \ion{S}{15} EW maps constructed with the linear EW method and the 2-D Spec EW method in Fig.~\ref{PaperIfig:Sabundance}. The \ion{S}{15} EW map constructed with the 2-D Spec EW method is quite noisy, probably because of the uncertainty in modeling the spectra at low signal-to-noise ratio. The \ion{S}{15} EW map constructed with the linear EW method better traces the S distribution, which is quite similar as the \ion{Si}{13} lines. The smoother appearance is apparently a result of the lower signal-to-noise ratio of S lines.

\subsubsection{Fe}\label{PaperIsubsubsec:metal_FeK}

Type~Ia SNRs are often thought to be the major source of Fe in the hot gas in and around galaxies (e.g., \citealt{Humphrey06,Ji09,Li09,Li13,Li15}). However, as a typical Type~Ia SNR, SN1006 is well known to exhibit a ``missing iron'' problem, i.e., has no significant Fe emission lines in most parts of the SNR (e.g., \citealt{Vink03}). Nevertheless, the existence of Fe is still revealed in some different ways. For example, \citet{Yamaguchi08} and \citet{Uchida13} have recently reported the detection of Fe~K emission lines (at a very low ionization state of $n_et<10^9\rm~cm^{-3}s$ and center at $\sim6.4\rm~keV$) with deep \emph{Suzaku} observations. On the other hand, optical and UV observations of bright background sources toward the direction of SN1006 have revealed some Fe absorption lines at low ionization states \citep{Fesen88,Wu93,Blair96,Hamilton97,Winkler05}. Therefore, it is most likely that the Fe-rich layers of SN1006 has not yet been completely shock heated by the reverse shock. In fact, these UV/optical spectra reveal that the reverse shock is penetrating still in a Si-rich zone \citep{Winkler05,Hamilton07}.

\begin{figure}[!h]
\begin{center}
\epsfig{figure=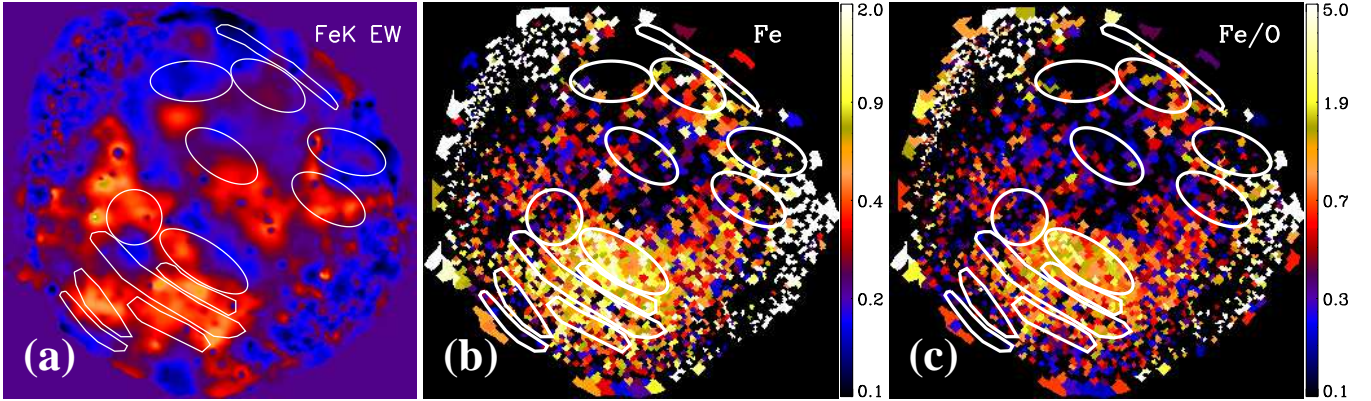,width=1.0\textwidth,angle=0, clip=}
\caption{(a) The Fe~K line EW map constructed with the linear EW method. (b) Fe abundance map. (c) Fe/O abundance ratio map. We caution that the Fe abundance is mainly determined with the widely distributed Fe~L lines at $\sim1\rm~keV$, different from the Fe~K line shown in panel~(a). The Fe~L lines typical do not appear as a prominent bump. We therefore do not construct EW maps for them.}\label{PaperIfig:Feabundance}
\end{center}
\end{figure}

Spatial distribution of Fe is shown in Fig.~\ref{PaperIfig:Feabundance}. The Fe~K-shell lines are very weak and only marginally detectable in our \emph{XMM-Newton} spectra extracted from the entire remnant (Fig.~\ref{PaperIfig:EWband}). In addition, the thermal continuum is mainly determined in soft X-ray ($\lesssim2\rm~keV$) and is thus poorly constrained in the Fe~K line band ($\sim6.2-6.7\rm~keV$). Therefore, we only present the Fe~K-line EW maps constructed with the linear EW method in Fig.~\ref{PaperIfig:Feabundance}a, together with the Fe abundance maps (Fig.~\ref{PaperIfig:Feabundance}b,c) mainly determined with the Fe~L-shell lines at $\sim1\rm~keV$. It is obvious that the strongest Fe emission appears in the SE quarter, consistent with the narrow band image and the spatially resolved spectral analysis conducted with the \emph{Suzaku} observations \citep{Yamaguchi08,Uchida13}.

It is interesting that the Fe abundance distribution (Fig.~\ref{PaperIfig:Feabundance}b,c) is quite similar as the \ion{O}{7}~$\rm K\delta-\zeta$/\ion{O}{8} EW ratio (Fig.~\ref{PaperIfig:Oabundance}h). The peak around the ``O Hole'' is so sharp that the \ion{O}{7}~$\rm K\delta-\zeta$/\ion{O}{8} ratio is at least twice lower and there is negligible Fe emission in other regions. The presence of the 6.4~keV Fe~K shell lines, as well as the relatively strong \ion{O}{7}~$\rm K\delta-\zeta$ lines, indicate that the ionization parameter should be low around the ``O Hole''. However, the measured $n_et$ peaks at this region (Fig.~\ref{PaperIfig:2DSpec_paraimg}b) apparently lead to opposite conclusions. Therefore, there are at least two components of the plasma toward the ``O Hole'', one with low $n_et$ of $\lesssim10^9\rm~cm^{-3}s$ (\citealt{Yamaguchi08,Uchida13}) accounting for the Fe~K and \ion{O}{7}~$\rm K\delta-\zeta$ emission lines, the other with high $n_et$ of $\sim5\times10^9\rm~cm^{-3}s$ accounting for most of the soft X-ray emission including the \ion{O}{7} and \ion{O}{8} lines. The high-$n_et$ component is most likely from the shocked ISM (e.g., \citealt{Ferrand12}), which is also consistent with the roughly solar O abundance. On the other hand, the low-$n_et$ component is probably the reserve shocked ejecta, which also explains the high Fe abundance. In the Type~Ia SNR SN1006, only the central region shows significant Fe emission (from both K-shell and L-shell transitions), which means the reverse shock has just reached the Fe-rich shell of the ejecta after $\gtrsim10^3\rm~yr$ since the SN explosion. 

\citet{Uchida13} suggested another possibility to explain the spectral fitting residual at 0.7-0.75~keV. There could be relatively strong Fe~L~$3d\rightarrow2p$ transitions of extremely low-ionization Fe ions (the most dominant charge state of Fe is less than Fe$^{16+}$), in addition to the \ion{O}{7}~$\rm K\delta-\zeta$ lines. This possibility is also suggested by \citet{Warren04} for the Type~Ia SNR E0509-67.5. Such contributions from newly shocked low Fe ions are also evidenced by the elevated emission at $\sim$0.8-0.9~keV of ``SE Shell 01'' (Fig.~\ref{PaperIfig:specindividual}d,h), which has similar properties as ``SE Shell 02 and 03'' but has significantly higher Fe abundance (Table~\ref{PaperItable:ParaIndividualSpec}, Fig.~\ref{PaperIfig:Feabundance}b,c). The Fe~L~$3d\rightarrow2p$ transitions of low Fe ions naturally explains the spatial coincidence of the 0.7-0.75~keV residual and the Fe abundance.\\

Many geometric models of SN1006 have been proposed in the literature, in order to explain the asymmetric distribution of multi-wavelength features, primarily based on UV/optical absorption line studies (e.g., \citealt{Hamilton97,Winkler05}) or X-ray mapping of prominent emission lines (e.g., \citealt{Uchida13}). In the above sections, we for the first time present maps of many physical parameters of SN1006, such as $kT$, $n_e$, $n_et$, and $t_{ion}$. We also present abundance and EW maps of many heavy elements. Most of these parameter maps show clear asymmetric distributions, which could be attributed to the asymmetric distributions of either the ambient ISM or the SN ejecta. For example, the Ne emission lines are very strong in the NW shell, consistent with an enhanced ISM density in this region as revealed in \ion{H}{1} and H$\alpha$ observations (e.g., \citealt{Dubner02,Nikolic13}; also see the electron density map shown in Fig.~\ref{PaperIfig:2DSpec_paraimg}d). However, the very different distributions of O, Ne, and Mg emission lines, which are usually thought to be good tracers of shocked ISM, indicate that the asymmetric distribution of ISM cannot explain all the characteristics of the spatial distribution of metals. The very asymmetric distribution of Si indicates that the reverse shock encounters the ejecta earlier in the SE half of the SNR. The detection of shocked Fe only in a small region further indicates the reverse shock reaches the Fe-rich ejecta very recently. These strong ejecta emission in the SE half is \emph{not} expected if the NW half has a higher surrounding ISM density and the SN exploded symmetrically in the center of the SNR. Therefore, it is very likely that the SN explosion is also intrinsically asymmetric, as expected theoretically (e.g., \citealt{GarciaSenz95}) and also suggested in many previous observational works (e.g., \citealt{Hamilton97,Uchida13}). 

\section{Summary and conclusions}\label{PaperIsec:Summary}

We develop new methods to conduct spatially resolved spectroscopy analysis in tessellated meshes each with total counts number above a given threshold. We apply this method to our \emph{XMM-Newton} LP and archival \emph{XMM-Newton} observations of SN1006, with total effective exposure times of 683, 710, and 439~ks of MOS-1, MOS-2, and PN. We extract spectra from 3596 tessellated regions each with a 0.3-8~keV total counts number $\gtrsim 10^4$. We fit these spectra with a simple ``VNEI+SRCUT'' model plus Gaussian lines representing \ion{O}{7}~$\rm K\delta-\zeta$ lines and sky background components. This simple spectral model with a 1-T thermal plasma is generally reliable to decompose the thermal and non-thermal emission and to characterize the average properties across the SNR. For the first time, we map out multiple physical parameters of the thermal plasma and the non-thermal component directly obtained from or derived based on the spectral fitting results. In particular, we adopt the shell-like geometric model introduced in \citet{Miceli12} and obtain an electron density map consistent with multi-wavelength estimates of the ambient ISM density.

We also develop a new method to construct equivalent width (EW) maps based on the continuum interpolation with the spectral model of each tessellated regions. This method has the advantages of accounting for the curvature of the underlying continuum and decomposing the thermal and non-thermal components. We compare the EW maps constructed with this method and the EW maps constructed with a common linear interpolation method as well as the abundance map from our spatially resolved spectral analysis. The spatial distributions of emission lines as revealed by these three methods are qualitatively consistent with each other. 

We further extract spectra from some larger regions (than the tessellated meshes) and compute average parameters in them to confirm the prominent features as revealed by the parameter and EW maps. These features are generally preserved, although sometimes less prominent because of the mixture of the spectra extracted from different meshes. We analyze these spectra extracted from the larger regions with the same 1-T model as adopted for individual meshes. The spectral analysis results are generally consistent with the parameter maps, except for some ``bad pixels'' caused by the inadequate modeling of a few isolated meshes.

In order to better characterize the average thermal and ionization states of SN1006, we construct probability distribution functions (PDFs) of $kT$ and $n_et$ of the thermal plasma. Previous measurements of $kT$ and $n_et$ in the literatures show large dispersions, but are all consistent with our measurements in different regions across the SNR. We thus conclude that the diversity of the $kT$ and $n_et$ measurements in the literatures is at least partially caused by the different selections of spectral regions. We characterize the $kT$ and $n_et$ PDFs with several Gaussian distributions. Only with such a modeling could we characterize the average thermal and ionization states of such an extended source, the gas in which spans a large range of hydrodynamical evolutionary stages. 

For the first time, we present maps across the entire SNR of $kT$, $n_et$, $t_{ion}$ (ionization age with time dimension), $n_e$, and metal abundances of the thermal component, as well as $\alpha$ (radio-to-X-ray slope) and $\nu_{cutoff}$ (cutoff frequency) of the non-thermal synchrotron emission. $\alpha$ is roughly constant in regions dominated by non-thermal emission, but $\nu_{cutoff}$ shows clearly spatial variations and is higher in brighter non-thermal filaments. In general, the outer shells have low $kT$, $n_et$, and $t_{ion}$, indicating that they are recently shocked ISM. The estimated electron number density represents a mixture of both shocked ISM and shocked SN ejecta, so could be adopted as upper limits of both components. The $n_e$ map follows the broad-band X-ray surface brightness distribution, except for the two non-thermal limbs. There is a ``Dark Belt'' presenting on the $n_e$ map, which also has higher Ne abundance, lower $n_et$, but comparable $t_{ion}$ than the surrounding regions. Based on the $n_e$ map, we could estimate the total mass of the X-ray emitting plasma under the shell like geometric model, which is $\sim14.5f^{\frac{1}{2}}\rm~M_\odot$. By comparing with the swept up ISM mass and the expected SN ejecta mass, we could further roughly constrain the volume filling factor $f\sim0.1$.

Spatial distribution of metals are traced with both the abundance maps and the EW maps constructed with either the linear EW method or the 2-D Spec EW method. O abundance is not significantly different from solar value over the entire remnant, and shows a low abundance hole roughly in the center of the SNR. The \ion{O}{7}~$\rm K\delta-\zeta$/\ion{O}{8} EW ratio is significantly higher in this ``O Hole'' than any other regions. This ``O Hole'' and the surrounding regions are also the only part of the SNR with significant Fe emission, from both low-ionization Fe~K lines and Fe~L lines. All these features indicate that the ``O Hole'' should be an Fe-rich ejecta shell recently reversed shocked. However, the high $n_et$ of this region also indicates that it should be largely comprised of shocked ISM. We therefore need a multi-temperature model to better decompose different thermal components. The ``NW Shell'', which is known to be blast wave shocked high density ISM, shows significantly enhanced Ne emission (but the Ne abundance is still subsolar), but just moderate O and Mg emissions. We thus conclude that Ne is mostly ISM in origin, but O and Mg may be partly from the ejecta. Si and S emissions are stronger in the SE half of SN1006, and is clearly different from O, Ne, Mg, and Fe emissions in spatial distribution. The asymmetric distributions of metals strongly suggest that there is either an asymmetric explosion of the SN or the ISM distribution is highly asymmetric around SN1006.

\acknowledgements

The authors would like to acknowledge Dyer Kristy and Reynolds Stephen for providing their radio images of SN1006 and for helpful discussions. We also acknowledge Ballet Jean and Acero Fabio for helpful discussions and the anonymous referee for helpful comments and suggestions. Li Jiang-Tao acknowledges the financial support from CNES and the travel support from NSFC through the grant 11233001.

{

}

\appendix

\section{Analysis of the sky background}\label{PaperIAppendix:SkyBackground}

After removing all the bright point-like sources, we extract the spectra of sky background in an annulus with radii of $20^\prime-25^\prime$ and centered at ($\alpha_{\rm J2000}=15^h02^m54.03^s$, $\delta_{\rm J2000}=-41^\circ55^\prime39.86^{\prime\prime}$) (Fig.~\ref{PaperIfig:tricolorregions}). We fit the spectra with a ``VMEKAL+power law'' model subjected to the foreground absorption plus a second MEKAL model with solar abundance and no foreground absorption. Here the VMEKAL and MEKAL models represent thermal plasma in an equilibrium ionization state with or without relative abundances set. The first thermal component (VMEKAL) characterize the combined contribution from the possibly scattered light from the remnant as well as the background radiation from the Galactic corona. The abundances of some key elements, for example, O, Ne, Fe, are set free in order to account for some prominent emission line features. The power law component, on the other hand, is used to characterize the possible contribution from the scattered light of the non-thermal arcs and the cosmic background mainly consists of unresolved active galactic nuclei (AGN) (e.g., \citealt{Moretti03}). The second thermal plasma component (MEKAL) represents the contribution from the local hot bubble, which is a structure in the solar neighbourhood, and is assumed to have solar abundance and no foreground absorption. The temperature of this component is fixed at 0.1~keV \citep{Kuntz00}. Such a background analysis is not aimed at well describing the sky background physically, but at roughly characterizing different components of the sky background for subtraction from the source spectra. A similar double subtraction procedure is often used in the study of extended sources (e.g., \citealt{Li08}).

\begin{figure}[!h]
\begin{center}
\epsfig{figure=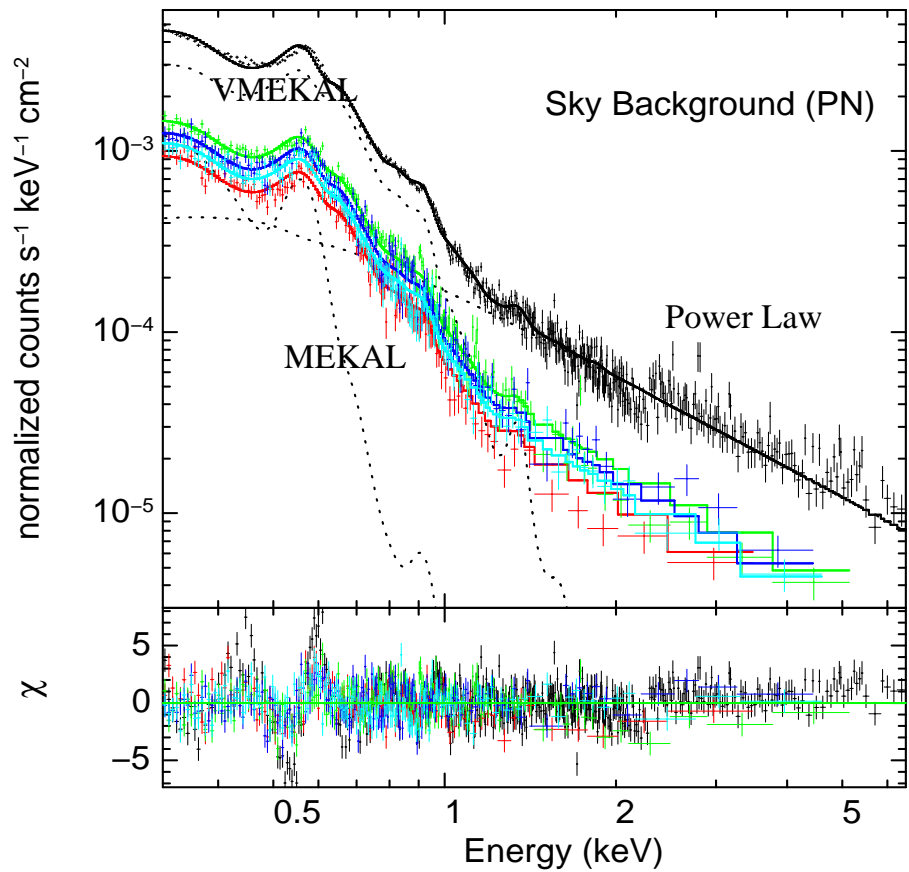,width=0.7\textwidth,angle=0, clip=}
\caption{Sky background spectrum (only PN to clarify) extracted from the whole annulus (black) and the four quadrants (colored) in Fig.~\ref{PaperIfig:tricolorregions}, with point-like sources subtracted. The dotted lines show different model components of the whole annulus as described in the text, while the solid line is a combination of all these components.}\label{PaperIfig:bkgspec}
\end{center}
\end{figure}

Fitting the background spectra with the above model results in a temperature of the VMEKAL component of $\sim0.2\rm~keV$ and subsolar abundances. These are consistent with the VMEKAL component of mainly Galactic halo in origin (e.g., \citealt{Lumb02}) and a negligible contribution from the scattered light of the SNR. The power law component has a photon index of $\sim1.4-1.7$ (slightly different for different instruments), significantly smaller than that of the synchrotron emission from the non-thermal arcs (e.g., \citealt{Allen08}), but roughly consistent with primarily cosmic in origin (from distant AGN) in both photon index and intensity (Fig.~\ref{PaperIfig:bkgspec}; \citealt{Lumb02}). This also indicates a negligible contribution from the scattered light of the non-thermal arcs of the SNR.

The above spectral analysis indicates a negligible contribution of the scattered light of the SNR to the sky background. The reason for this low scattering is that the foreground \ion{H}{1} (and so dust) column density toward SN1006 is low ($N_{\rm H}=6.8\times10^{20}\rm~cm^{-2}$; \citealt{Dubner02}), given that SN1006 lies high above the Galactic plane. We further examine the azimuthal variation of the sky background to confirm this conclusion. X-ray emission of SN1006 shows significant azimuthal variation in both spectral shape and total intensity. Therefore, the sky background may have non-negligible azimuthal variations if the scattered light is important. We examine this possibility by extracting background spectra from each of the four quadrants of the annulus as shown in Fig.~\ref{PaperIfig:tricolorregions}, with azimuthal angles of $13^\circ-103^\circ$, $103^\circ-193^\circ$, $193^\circ-283^\circ$, and $283^\circ-13^\circ$ from west. As shown in Fig.~\ref{PaperIfig:bkgspec}, the flux variation in different quadrants is small, and could be largely attributed to the variation of the effective area scale of different regions. The scatter of the intrinsic intensity of the sky background is $\sim8\%$ (at $1~\sigma$ confidence level). Since the sky background only contributes a small fraction of the flux in the source regions (e.g., Fig.~\ref{PaperIfig:spec_mesh}), this fluctuation in background flux should have negligible effect on spectral modeling. Furthermore, we do not find any significant spectral variation in the four quadrants. In fact, the models of the four quadrants shown in Fig.~\ref{PaperIfig:bkgspec} are exactly the same as that of the integrated spectra of the whole annulus, except for a constant normalization factor multiplied to all the model components. In conclusion, it is safe to apply the sky background from the whole annulus to all the source spectra.


\begin{thebibliography}

\scriptsize

\bibitem[Acero et al.(2007)]{Acero07} Acero F., Ballet J., Decourchelle A., 2007, A\&A, 475, 883

\bibitem[Acero et al.(2010)]{Acero10} Acero F. et al., 2010, A\&A, 516, 62

\bibitem[Acero et al.(2015)]{Acero15} Acero F., Lemoine-Goumard M., Renaud M., Ballet J., Hewitt J. W., Rousseau R., Tanaka, T., 2015, astro-ph$\backslash$arXiv:1506.02307

\bibitem[Aharonian et al.(2005)]{Aharonian05} Aharonian F. et al., 2005, A\&A, 437, 135

\bibitem[Allen et al.(2001)]{Allen01} Allen G. E., Petre R., Gotthelf E. V., 2001, ApJ, 558, 739

\bibitem[Allen et al.(2008)]{Allen08} Allen G. E., Houck J. C., Sturner S. J., 2008, ApJ, 683, 773

\bibitem[Becker et al.(1980)]{Becker80} Becker R. H., Szymkowiak A. E., Boldt E. A., Holt S. S., Serlemitsos P. J., 1980, ApJ, 240, 33

\bibitem[Blair et al.(1996)]{Blair96} Blair W. P., Long K. S., Raymond J. C., 1996, ApJ, 468, 871

\bibitem[Borkowski et al.(2001)]{Borkowski01} Borkowski K. J., Lyerly W. J., Reynolds S. P., 2001, ApJ, 548, 820

\bibitem[Broersen et al.(2013)]{Broersen13} Broersen S., Vink J., Miceli M., Bocchino F., Maurin G., Decourchelle A., 2013, A\&A, 552, 9

\bibitem[Broos et al.(2010)]{Broos10} Broos P. S., Townsley L. K., Feigelson E. D., Getman K. V., Bauer F. E., Garmire G. P., 2010, ApJ, 714, 1582

\bibitem[Burleigh et al.(2000)]{Burleigh00} Burleigh M. R., Heber U., O'Donoghue D., Barstow M. A., 2000, A\&A, 356, 585

\bibitem[Cappellari \& Copin(2003)]{Cappellari03} Cappellari M., Copin Y., 2003, MNRAS, 342, 345

\bibitem[Cassam-Chena\"{\i} et al.(2004)]{CassamChenai04} Cassam-Chena\"{\i} G., Decourchelle A., Ballet J., Sauvageot J.-L., Dubner G., Giacani E., 2004, A\&A, 427, 199

\bibitem[Cassam-Chena\"{\i} et al.(2008)]{CassamChenai08} Cassam-Chena\"{\i} G., Hughes J. P., Reynoso E. M., Badenes C., Moffett D., 2008, ApJ, 680, 1180

\bibitem[Chen et al.(2008)]{Chen08} Chen Y., Seward F. D., Sun M., Li J.-T., 2008, ApJ, 676, 1040

\bibitem[Decourchelle et al.(2000)]{Decourchelle00} Decourchelle A., Ellison D. C., Ballet J., 2000, ApJL, 543, 57

\bibitem[Dennerl et al.(2004)]{Dennerl04} Dennerl K. et al., 2004, SPIE, 5488, 61

\bibitem[Diehl \& Statler(2006)]{Diehl06} Diehl S., Statler T. S., 2006, MNRAS, 368, 497

\bibitem[Dubner et al.(2002)]{Dubner02} Dubner G. M., Giacani E. B., Goss W. M., Green A. J., Nyman L.-A., 2002, A\&A, 387, 1047

\bibitem[Dyer et al.(2001)]{Dyer01} Dyer K. K., Reynolds S. P., Borkowski K. J., Allen G. E., Petre R., 2001, ApJ, 551, 439

\bibitem[Dyer et al.(2009)]{Dyer09} Dyer K. K., Cornwell T. J., Maddalena R. J., 2009, AJ, 137, 2956

\bibitem[Ferrand et al.(2012)]{Ferrand12} Ferrand G., Decourchelle A., Safi-Harb S., 2012, ApJ, 760, 34

\bibitem[Fesen et al.(1988)]{Fesen88} Fesen R. A., Wu C.-C., Leventhal M., Hamilton A. J. S., 1988, ApJ, 327, 164

\bibitem[Foster et al.(2012)]{Foster12} Foster A. R., Ji L., Smith R. K., Brickhouse N. S., 2012, ApJ, 756, 128

\bibitem[Gaensler(1998)]{Gaensler98} Gaensler B. M., 1998, ApJ, 493, 781

\bibitem[Garcia-Senz \& Woosley(1995)]{GarciaSenz95} Garcia-Senz D., Woosley S. E., 1995, ApJ, 454, 895

\bibitem[Ghavamian et al.(2002)]{Ghavamian02} Ghavamian P., Winkler P. F., Raymond J. C., Long K. S., 2002, ApJ, 572, 888

\bibitem[Hamilton et al.(1997)]{Hamilton97} Hamilton A. J. S., Fesen R. A., Wu C. C., Crenshaw D. M., Sarazin C. L., 1997, ApJ, 482, 838

\bibitem[Hamilton et al.(2007)]{Hamilton07} Hamilton A. J. S., Fesen R. A., Blair W. P., 2007, MNRAS, 381, 771

\bibitem[Heng et al.(2007)]{Heng07} Heng K., van Adelsberg M., McCray R., Raymond J. C., 2007, ApJ, 668, 275

\bibitem[Hodges-Kluck \& Reynolds(2012)]{HodgesKluck12} Hodges-Kluck E. J., Reynolds C. S., 2012, ApJ, 746, 167

\bibitem[Humphrey \& Buote(2006)]{Humphrey06} Humphrey P. J., Buote D. A. 2006, ApJ, 639, 136

\bibitem[Hwang et al.(2000)]{Hwang00} Hwang U., Holt S. S., Petre R., 2000, ApJL, 537, 119

\bibitem[Iwamoto et al.(1999)]{Iwamoto99} Iwamoto K., Brachwitz F., Nomoto K., Kishimoto N., Umeda H., Hix W. R., Thielemann F.-K., 1999, ApJS, 125, 439

\bibitem[Ji et al.(2009)]{Ji09} Ji J., Irwin J. A., Athey A., Bregman J. N., Lloyd-Davies E. J. 2009, ApJ, 696, 2252

\bibitem[Jones \& Pye(1989)]{Jones89} Jones L. R., Pye J. P., 1989, MNRAS, 238, 567

\bibitem[Katsuda et al.(2010)]{Katsuda10} Katsuda S., Petre R., Mori K., Reynolds S. P., Long K. S., Winkler P. F., Tsunemi H., 2010, ApJ, 723, 383

\bibitem[Kirshner et al.(1987)]{Kirshner87} Kirshner R., Winkler P. F., Chevalier R. A., 1987, ApJ, 315, 135

\bibitem[Korreck et al.(2004)]{Korreck04} Korreck K. E., Raymond J. C., Zurbuchen T. H., Ghavamian P., 2004, ApJ, 615, 280

\bibitem[Koyama et al.(1995)]{Koyama95} Koyama K., Petre R., Gotthelf E. V., Hwang U., Matsuura M., Ozaki M., Holt S. S., 1995, Nature, 378, 255

\bibitem[Koyama et al.(2008)]{Koyama08} Koyama K., Yamaguchi H., Bamba A., 2008, ChJAS, 8, 165

\bibitem[Kuntz \& Snowden(2000)]{Kuntz00} Kuntz K. D., Snowden S. L., 2000, ApJ, 543, 195

\bibitem[Kuntz \& Snowden(2008)]{Kuntz08} Kuntz K. D., Snowden S. L., 2008, A\&A, 478, 575

\bibitem[Laming et al.(1996)]{Laming96} Laming J. M., Raymond J. C., McLaughlin B. M., Blair W. P., 1996, ApJ, 472, 267

\bibitem[Leahy et al.(1991)]{Leahy91} Leahy D. A., Nousek J., Hamilton A. J. S., 1991, ApJ, 374, 218

\bibitem[Li et al.(2008)]{Li08} Li J.-T., Li Z., Wang Q. D., Irwin J. A., Rossa J., 2008, MNRAS, 390, 59

\bibitem[Li et al.(2009)]{Li09} Li J.-T., Wang Q. D., Li Z., Chen Y., 2009, ApJ, 706, 693

\bibitem[Li \& Wang(2013)]{Li13} Li J.-T., Wang Q. D., 2013, MNRAS, 435, 3071

\bibitem[Li (2015)]{Li15} Li J.-T., 2015, MNRAS, in press, astro-ph$\backslash$arXiv:1507.07040

\bibitem[Long et al.(1988)]{Long88} Long K. S., Blair W. P., van den Bergh S., 1988, ApJ, 333, 749

\bibitem[Long et al.(2003)]{Long03} Long K. S., Reynolds S. P., Raymond J. C., Winkler P. F., Dyer K. K., Petre R., 2003, ApJ, 586, 1162

\bibitem[Lopez et al.(2013)]{Lopez13} Lopez L. A., Ramirez-Ruiz E., Castro D., Pearson S., 2013, ApJ, 764, 50

\bibitem[Lu \& Aschenbach(2000)]{Lu00} Lu F. J., Aschenbach B., 2000, A\&A, 362, 1083

\bibitem[Lumb et al.(2002)]{Lumb02} Lumb D. H., Warwick R. S., Page M., De Luca A., 2002, A\&A, 389, 93

\bibitem[Miceli et al.(2009)]{Miceli09} Miceli M., et al., 2009, A\&A, 501, 239

\bibitem[Miceli et al.(2012)]{Miceli12} Miceli M., Bocchino F., Decourchelle A., Maurin G., Vink J., Orlando S., Reale F., Broersen S., 2012, A\&A, 546, 66

\bibitem[Miceli et al.(2013)]{Miceli13} Miceli M., Bocchino F., Decourchelle A., Vink J., Broersen S., Orlando S., 2013, A\&A, 556, 80

\bibitem[Miceli et al.(2014)]{Miceli14} Miceli M., Acero F., Dubner G., Decourchelle A., Orlando S., Bocchino F., 2014, ApJL, 782, 33

\bibitem[Moffett et al.(1993)]{Moffett93} Moffett D. A., Goss W. M., Reynolds S. P., 1993, AJ, 106, 1566

\bibitem[Moretti et al.(2003)]{Moretti03} Moretti A., Campana S., Lazzati D., Tagliaferri G., 2003, ApJ, 588, 696

\bibitem[Nikoli$\rm\acute{c}$ et al.(2013)]{Nikolic13} Nikoli$\rm\acute{c}$ S., van de Ven G., Heng K., Kupko D., Husemann B., Raymond J. C., Hughes J. P., Falc$\rm\acute{o}$n-Barroso J., 2013, Science, 340, 45

\bibitem[Pye et al.(1981)]{Pye81} Pye J. P., Pounds K. A., Rolf D. P., Smith A., Willingale R., Seward F. D., 1981, MNRAS, 194, 569

\bibitem[Randall et al.(2008)]{Randall08} Randall S., Nulsen P., Forman W. R., Jones C., Machacek M., Murray S. S., Maughan B., 2008, ApJ, 688, 208

\bibitem[Raymond et al.(2007)]{Raymond07} Raymond J. C., Korreck K. E., Sedlacek Q. C., Blair W. P., Ghavamian P., Sankrit R., 2007, ApJ, 659, 1257

\bibitem[Ressler et al.(2014)]{Ressler14} Ressler S. M., Katsuda S., Reynolds S. P., Long K. S., Petre R., Williams B. J., Winkler P. F., 2014, ApJ, 790, 85

\bibitem[Reynolds \& Gilmore(1986)]{Reynolds86} Reynolds S. P., Gilmore D. M., 1986, AJ, 92, 1138

\bibitem[Reynolds \& Keohane(1999)]{Reynolds99} Reynolds, S. P., Keohane, J. W., 1999, ApJ, 525, 368

\bibitem[Reynolds(2008)]{Reynolds08} Reynolds S. P., 2008, ARA\&A, 46, 89

\bibitem[Rothenflug et al.(2004)]{Rothenflug04} Rothenflug R., Ballet J., Dubner G., Giacani E., Decourchelle A., Ferrando P., 2004, A\&A, 425, 121

\bibitem[Slane(2014)]{Slane14} Slane P., 2014, IAUS, 296, 226

\bibitem[Stephenson(2010)]{Stephenson10} Stephenson F. R. 2010, A\&G, 51, 27

\bibitem[Uchida et al.(2013)]{Uchida13} Uchida H., Yamaguchi H., Koyama K., 2013, ApJ, 771, 56

\bibitem[Vink et al.(2000)]{Vink00} Vink J., Kaastra J. S., Bleeker J. A. M., Preite-Martinez A., 2000, A\&A, 354, 931

\bibitem[Vink et al.(2003)]{Vink03} Vink J., Laming J. M., Gu M. F., Rasmussen A., Kaastra J. S., 2003, ApJL, 587, 31

\bibitem[Vink et al.(2010)]{Vink10} Vink J., Yamazaki R., Helder E. A., Schure K. M. 2010, ApJ, 722, 1727

\bibitem[Vink(2012)]{Vink12} Vink J., 2012, A\&ARv, 20, 49

\bibitem[Warren \& Hughes(2004)]{Warren04} Warren J. S., Hughes J. P., 2004, ApJ, 608, 261

\bibitem[Winkler \& Long(1997)]{Winkler97} Winkler P. F., Long K. S., 1997, ApJ, 491, 829

\bibitem[Winkler et al.(2003)]{Winkler03} Winkler P. F., Gupta G., Long K. S., 2003, ApJ, 585, 324

\bibitem[Winkler et al.(2005)]{Winkler05} Winkler P. F., Long K. S., Hamilton A. J. S., Fesen R. A., 2005, ApJ, 624, 189

\bibitem[Winkler et al.(2013)]{Winkler13} Winkler P. F., Williams B. J., Blair W. P., Borkowski K. J., Ghavamian P., Long K. S., Raymond J. C., Reynolds S. P., 2013, ApJ, 764, 156

\bibitem[Winkler et al.(2014)]{Winkler14} Winkler P. F., Williams B. J., Reynolds S. P., Petre R., Long K. S., Katsuda S., Hwang U., 2014, ApJ, 781, 65

\bibitem[Wu et al.(1993)]{Wu93} Wu C.-C., Crenshaw D. M., Fesen R. A., Hamilton A. J. S., Sarazin C. L., 1993, ApJ, 416, 247

\bibitem[Yamaguchi et al.(2008)]{Yamaguchi08} Yamaguchi H. et al., 2008, PASJ, 60, 141

\end{thebibliography}
\end{document}